\pdfoutput=1
\documentclass[10pt,a4paper]{article}
\usepackage{jheppub,graphicx,hyperref,multirow,subfig}
\usepackage[dvipsnames]{xcolor}
\usepackage[utf8]{inputenc}
\usepackage{listings}
\hypersetup{colorlinks=true, linkcolor=Maroon, citecolor=Maroon, filecolor=Maroon, urlcolor=Maroon}

\newcommand{\be}{\begin{equation}}
\newcommand{\ee}{\end{equation}}
\def\bsp#1\esp{\begin{split}#1\end{split}}

\newcommand{\eg}{\textit{e.g.}}

\newcommand{\ds}{\displaystyle }

\newcommand{\amc}{{\sc MadGraph5}\_a{\sc MC@NLO}}
\newcommand{\fr}{{\sc Feyn\-Rules}}
\newcommand{\fa}{{\sc Feyn\-Arts}}

\newcommand{\ma}{{\sc MadAnalysis~5}}
\newcommand{\py}{{\sc Pythia~8}}
\newcommand{\del}{{\sc Delphes~3}}
\newcommand{\fj}{{\sc FastJet}}

\newcommand{\pmerr}[2]{^{+#1\%}_{-#2\%}}

\def\fa{t_{i}}

\def\fab{\bar{t}_{j}}

\def\Sop{\bar{t} t }
\def\Pop{\bar{t}  \gamma^5 t }
\def\Vop{\bar{t} \gamma t }
\def\Aop{\bar{t} \gamma \gamma^5 t }
\def\Top{\bar{t} \sigma  t }
\def\DTop{\bar{t} \sigma  \gamma^5 t }

\def\oRR{\scr{O}^1_{RR}}
\def\oLL{\scr{O}^1_{LL}}
\def\oLR{\scr{O}^1_{LR}}
\def\oTaLR{\scr{O}^8_{LR}}
\def\oTaLL{\scr{O}^8_{LL}}
\def\os{\scr{O}^1_{S}}
\def\oTas{\scr{O}^8_{S}}

\def\ops{\scr{O}^1_{PS}}
\def\oTaps{\scr{O}^8_{PS}}

\def\ysS{y_{1S}}
\def\ysP{y_{1P}}
\def\yoS{y_{8S}}
\def\yoP{y_{8P}}

\def\gsL{g_{1L}}
\def\gsR{g_{1R}}
\def\goL{g_{8L}}
\def\goR{g_{8R}}

\def\tphi{X_t}
\def\mxt{m_{X_t}}
\def\tphi{X}
\def\mxt{m_{X}}

\def\sing{S_1}
\def\oct{S_8}

\def\vs{V_1{}}
\def\vo{V_8{}}

\def\np{{\rm NP}}
\def\nptwo{{\rm NP}^2{}}
\def\Int{{\rm Int}}

\newcommand{\eNPij}[1]{\varepsilon_{\rm{NP}^2,#1}}
\newcommand{\eInti}[1]{\varepsilon_{\rm Int,#1}}

\newcommand{\scr}[1]{\ensuremath{\mathcal{#1}}}

\newcommand*{\CPthree}{~\small \textit{Centre  for  Cosmology,  Particle  Physics  and  Phenomenology  (CP3), Universitée  catholique  de  Louvain,  1348  Louvain-la-Neuve,  Belgium}}
\newcommand*{\BOLOGNA}{~\small  \textit{Dipartimento  di  Fisica  e  Astronomia,  Universit\`a  di  Bologna  e  INFN, Sezione  di  Bologna,  via  Irnerio  46,  I-40126  Bologna,  Italy}}
\newcommand*{\INFNFR}{~\small \textit{Istituto Nazionale di Fisica Nucleare, Laboratori Nazionali di Frascati, C.P. 13, 00044 Frascati, Italy}}
\newcommand*{\LPTHE}{~\small \textit{Sorbonne Universit\'e, CNRS, Laboratoire de Physique Théorique et Haute \'Energies, LPTHE, F-75005 Paris France}}
\newcommand*{\UNI}{~\small  \textit{Institut  Universitaire  de  France,  103  boulevard  Saint-Michel,  75005  Paris,  France}}

\author[a]{Luc~Darmé}
\author[b,c]{\!\!, Benjamin~Fuks}
\author[d,e]{\!, and Fabio~Maltoni}
\emailAdd{luc.darme@lnf.infn.it}
\emailAdd{fuks@lpthe.jussieu.fr} 
\emailAdd{fabio.maltoni@unibo.it} 

\affiliation[a]{\INFNFR}
\affiliation[b]{\LPTHE}
\affiliation[c]{\UNI}
\affiliation[d]{\CPthree}
\affiliation[e]{\BOLOGNA}

\title{Top-philic heavy resonances in four-top final states and their EFT interpretation}

\date{\today}

\abstract{
	With an expected rate of about one event per 100,000 top-quark pairs, four top-quark final states very rarely arise at the LHC. Though scarce, they offer a unique window onto top-quark compositeness, self-interactions and more generically, onto any top-philic new physics. By employing  simplified models featuring heavy resonances, we study the range of validity of effective theory interpretations of current four top-quark analyses at the LHC and establish their future reach at the HL-LHC. We find that for the class of models under consideration, the effective field theory interpretations are not applicable. We therefore present the most up-to-date limits obtained from public CMS analyses using simplified models. Finally, we put forward a novel recasting strategy for the experimental results based on the production of top quarks with large transverse momentum.
}

\begin{document}
\maketitle
\flushbottom

\section{Introduction}
With the third run of the LHC at the horizon, both the CMS and ATLAS experiments will have soon access to about $400$ fb$^{-1}$ of data.  This large statistics opens up the possibility to search for rare phenomena, corresponding to  production cross sections of a few femtobarns. Of particular interest in this regard is the four-top production mode. While characterised by a small Standard Model (SM) cross section at 13 TeV, $\sigma_{4t}^{\rm SM} = 11.97^{+2.15}_{-2.51}~{\rm fb}$~\cite{Frederix:2017wme}, it leads to very energetic final states with a distinctive signature and a low background, leaving hence ample room to study new physics contributions. 

Four-top hadroproduction, $pp\to t
\bar{t}t\bar{t}$, proceeds mainly via pure QCD interactions. Electroweak (EW) contributions are relatively large, yet tend to cancel each other both at the Born level as well as at the next-to-leading-order (NLO) level~\cite{Frederix:2017wme}. In particular, the contributions from Higgs-boson exchange between two top-quark lines lead to a polynomial sensitivity of the cross section on the top-quark Yukawa coupling, which provides a complementary handle for its determination relative to the $t\bar tH$ channel~\cite{Sirunyan:2017roi,Aaboud:2018xpj,Sirunyan:2019wxt}. Moreover, the $pp\to t \bar{t}t\bar{t}$ process could also be mediated by new top-philic particles, as predicted by a variety of new physics scenarios~\cite{Battaglia:2010xq,Dev:2014yca,Greiner:2014qna,Alvarez:2016nrz,Kim:2016plm,Fox:2018ldq}. Examples of such new particles arise in composite Higgs solutions to the hierarchy problem~\cite{Lillie:2007hd,Pomarol:2008bh,Zhou:2012dz,Cacciapaglia:2015eqa,Liu:2019bua,Cacciapaglia:2020vyf}, in models based on minimal flavour violation~\cite{Gerbush:2007fe,Hayreter:2017wra}, in models featuring extended supersymmetries~\cite{Salam:1974xa,Fayet:1974pd,Fayet:1975yi,AlvarezGaume:1996mv,Fox:2002bu,Plehn:2008ae,Choi:2008ub,GoncalvesNetto:2012nt,Fuks:2012im,Calvet:2012rk,Benakli:2014cia,Beck:2015cga,Kotlarski:2016zhv,Darme:2018dvz,Carpenter:2020hyz,Carpenter:2020evo}, or simply when the particle spectrum includes additional scalar fields interacting with the Standard Model via their mixing with the Higgs boson. The phenomenology of light top-philic particles, {\it i.e.} with a mass below the top mass $m_t$, has been studied in the past~\cite{Alvarez:2019uxp}, and an emphasis has been put on the associated production of such particles with a top quark pair~\cite{Azevedo:2020fdl}. Several phenomenological studies have appeared recently with the goal to bound new physics contributions in four top-quark production in the Standard Model Effective Field Theory (SMEFT) framework~\cite{Agram:2013koa,Khatibi:2014via,Durieux:2014xla, Guo:2016kea,Shen:2018mlj,DHondt:2018cww,Hartland:2019bjb,Degrande:2020evl,Liu:2020bem,Banelli:2020iau}. While powerful, the SMEFT approach has also intrinsic limitations that could become relevant in the four-top production case.  First, when truncated at dimension six, the SMEFT does not describe all four-top interactions which could be potentially relevant in a new physics framework, such as those mediated by scalar singlets of $SU(2)_{\rm L}$. Second, in the LHC context where the sensitivity is limited to scales that are possibly of the same order of those naturally associated with the final states under consideration, the validity itself of such an approach needs to be scrutinised. 

In this work, we compare the phenomenology of a set of perturbative top-philic simplified models, including scalar and vector particles that could be either colour-singlet or colour-octet, with the one obtained when considering the corresponding Effective Field Theory (EFT) limit in an $SU(3)_{\rm C} \times U(1)_{\rm em}$-invariant way. We establish the relation between the SMEFT Warsaw basis~\cite{Grzadkowski:2010es,AguilarSaavedra:2010zi,AguilarSaavedra:2018nen} to the set of operators relevant for four-top production, and we match the latter operators at the tree level to the considered simplified models when taking the limit of heavy top-philic mediators. This allows for a direct comparison of the constraints derived from the results of the LHC in the simplified model approach, with those derived in the corresponding EFT limit. Our work is also motivated by recent studies~\cite{Banelli:2020iau,Alvarez:2020ffi} that suggest that tensions in ATLAS $t\bar th$ and $t\bar t t \bar t$ observations with respect to the SM predictions could be an indication of the existence of a top-philic new boson with a mass of ${\cal O}(100)$ GeV. 

As we expect new physics contributions to significantly affect the kinematics of four-top production relatively to the SM, we derive limits on the simplified model and the EFT by estimating the number of signal events expected to populate the different signal regions of standard four-top LHC analyses. In practice, we employ a chain of various high-energy physics programs to generate signal events ({\sc FeynRules}~\cite{Christensen:2009jx,Degrande:2011ua,Alloul:2013bka}, {\sc MadGraph5\_aMC@NLO}~\cite{Alwall:2014hca}, {\sc Pythia}~8~\cite{Sjostrand:2014zea} and {\sc Delphes}~3~\cite{deFavereau:2013fsa}) and implement a typical four-top experimental analysis (that we choose to be the CMS analysis of ref.~\cite{Sirunyan:2019wxt}) in the  \ma\ platform~\cite{Conte:2012fm,Conte:2014zja,Dumont:2014tja,Conte:2018vmg}. Details on the validation of this implementation have been published separately~\cite{Darme:2020hxc,Fuks:2021wpe}, so that  they will be only briefly summarised in this work. \ma\ predictions for the signal efficiencies are used to derive exclusion contours at the 95\% confidence level (C.L.) in the parameter space of top-philic models. In order to emphasise the need for dedicated new physics search strategies at the next LHC operation runs when targeting a four top-quark final state, we also present a simple high-$H_T$ interpretation of the CMS results. We find that the latter provides stronger constraints on simplified model scenarios, mostly when new top-philic particles can be produced on-shell and transfer a large transverse momentum to the top quarks originating from their decay.

The main result of this work is that the current LHC sensitivity to new physics through four-top probes, and to a good extent the future HL-LHC one, only covers the on-shell production of heavy top-philic resonances when all new couplings are perturbative. Using an EFT approach would then significantly underestimate the exclusions in almost all regions of the parameter spaces of the simplified models considered. As we will see, the signal efficiencies resulting from both the EFT and the simplified model approaches are nonetheless relatively similar in the case of the CMS analysis that we have reinterpreted, allowing for a simple cross-section rescaling. We further stress that our results do not cover the case of couplings entering the non-perturbative regime, as expected for instance in various composite Higgs scenarios.

This work is organised as follows. In Sec.~\ref{sec:theory} we present Lagrangian formulations for the considered simplified models that all feature a different class of new top-philic bosons, we construct the minimal set of four-top operators needed to match these models to an EFT framework and discuss four-top production at the LHC. In Sec.~\ref{sec:recast} we describe the experimental search used to constrain new physics-induced four-top production at the LHC, as well as our calculation and simulation setup. Finally, Sec.~\ref{sec:results} is dedicated to the presentation of our results. We show limits derived from CMS four-top data and therefore obtain the most up-to-date constraints on the considered top-philic simplified model scenarios. Prospects at the future high-luminosity LHC (HL-LHC) run are also estimated. We summarise our work and conclude in Sec.~\ref{sec:concl}.

\section{Simplified models and effective field theory}\label{sec:theory}

In this section we first introduce and motivate the simplified models on which the reinterpretation of the four-top searches performed in this study is based.  We then establish the connection of these models to an effective field theory featuring four-top contact interactions at the dimension-six level. Finally, we compare predictions for four-top production in the EFT and resonant scenarios, and discuss their matching.

\subsection{Model definitions}\label{sec:model}
We consider various simplified models relevant for beyond the Standard Model top-philic physics. To be as general as possible, we define those models after electroweak symmetry breaking, so that the associated Lagrangians are invariant under the $SU(3)_{\rm C} \times U(1)_{\rm em}$ gauge group. Such new physics scenarios naturally arise in various UV constructions where the new interactions of the top quark are generated after electroweak symmetry breaking (or after the breaking of a potentially larger symmetry). In these cases, when connecting the UV physics to the TeV-scale one, chirality flips in diagrams computed in the UV theory do not significantly suppress the corresponding low-energy effective couplings, as the latter are simply reduced by a factor of $m_t^2/v^2$ of ${\cal O}(1)$ (where $m_t$ stands for the top quark mass and $v$ for the vacuum expectation value of the SM Higgs field). This situation corresponds to what arises for instance in the usual Two-Higgs-Doublet models (2HDM) (see \eg\ refs.~\cite{Branco:2011iw,Dev:2014yca,Kraml:2019sis}). 

In this work, we account for new spin-0 ($S$) and spin-1 ($V$) states and we consider scenarios where these new states are either colour-singlets ($\sing$ and $\vs$) or colour-octets ($\oct$ and $\vo$). In the following, we will often use the generic notation $\tphi$ to refer to any of such  top-philic particles. While fields in the sextet representation of $SU(3)_{\rm C}$ or with a tensorial Lorentz structure could be treated similarly, such cases rarely arise in common UV setups, so that they are omitted from the present report. 

The new physics Lagrangian describing the dynamics of a top-philic colour-singlet real scalar state $\sing$ of mass $m_{\sing}$, and the one associated with the corresponding vector state $\vs$ of mass $m_{\vs}$, can be written as
\begin{align}\label{eq:Lss}
  \scr{L}_{\sing} = \frac{1}{2} \partial_\mu \sing \partial^\mu \sing- \frac{1}{2} m_{\sing}^2 \sing^2 + \bar{t} \left[ \ysS  + i\ \ysP \gamma^5 \right] \sing ~ \! t  \ ,
\end{align}
and 
\begin{align}\label{eq:Lvs}
 \scr{L}_{\vs} = - \frac{1}{4} \vs^{\mu \nu} \vs_{\mu \nu} + \frac{1}{2} m_{\vs}^2 \vs^\mu \vs_\mu +  \bar{t} \gamma_\mu \left[ \gsL P_L + \gsR P_R \right] \vs^\mu ~ \! t   \ ,
\end{align}
respectively. In general, the scalar and pseudo-scalar Yukawa couplings $\ysS$ and $\ysP$ appearing in eq.~\eqref{eq:Lss} can be complex parameters, in which case one needs to add Hermitian conjugate contributions to the above Lagrangian. In the limiting case where the $\sing$ state is of a definite scalar (pseudo-scalar) nature, then only the $\ysS$ ($\ysP$) coupling is non-zero and real. For simplicity, we only consider these $CP$-conserving scenarios that involve a single real new physics coupling. A typical example of a UV model whose Lagrangian after electroweak symmetry breaking includes the simplified model Lagrangian~\eqref{eq:Lss} could be the type-I or type-II 2HDM~\cite{Branco:2011iw,Dev:2014yca,Kraml:2019sis}. The vector case~\eqref{eq:Lvs}, involving the left-handed and right-handed coupling strengths $\gsL$ and $\gsR$ (taken, for simplicity, real and equal in the phenomenological analyses considered in this study), can be generated, for instance, in composite models that feature the mixing of the top-quark with new vector-like states and that include extra scalar, fermion and vector resonances~\cite{Franzosi:2016aoo}, or in non-minimal models motivated by the $B$-anomalies that feature leptoquarks, vector-like quarks and extra gauge bosons~\cite{Calibbi:2017qbu,Faber:2018qon}. 
While any UV-complete theory generally predicts various other signals at colliders, these will be ignored in the simplified model approach pursued in this work, and we will solely focus on the four-top quark signals originating from the above two Lagrangians.

We now turn to the second class of simplified models investigated in this work. Those models involve either a scalar state $\oct$ of mass $m_{\oct}$, or a vector state $\vo$ of mass $m_{\vo}$. This time, the two fields are both colour-octet with respect to $SU(3)_{\rm C}$. The new physics Lagrangian, for the case of the real scalar field $\oct$, reads
\begin{align}\label{eq:Lso}
  \scr{L}_{\oct} = \frac{1}{2} D_\mu \oct^A D^\mu \oct^A - \frac{1}{2} m_{\oct}^2 \oct^A \oct^A + \bar{t} \left[ \yoS +  i\ \yoP \gamma^5 \right] T^A \oct^A ~ \! t   \ .
\end{align}
In this expression, we have introduced  the generators of $SU(3)$ in the fundamental representation $T^A$,
we have ignored any higher-dimensional operator coupling the scalar-octet state to gluons and we have included the scalar ($\yoS$) and pseudo-scalar ($\yoP$) interactions of the $\oct$ state with the top quark. In addition, the usual QCD interactions of the scalar field are included through the covariant derivatives used for the kinetic term. Similarly to the scalar-singlet case, the new physics couplings are in general complex. The interesting limiting cases of a purely scalar (pseudo-scalar) $\oct$ field, that will be examined in this work, are recovered by setting $\yoP$ ($\yoS$) to zero, $\yoS$ ($\yoP$) being real. These configurations find their motivation, for instance, in non-minimal supersymmetric models that feature scalar and pseudo-scalar sgluon fields~\cite{Salam:1974xa,Fayet:1974pd,Fayet:1975yi,AlvarezGaume:1996mv}.

The Lagrangian describing the dynamics of the vector-octet state $\vo$ of mass $m_{\vo}$ is given by
\begin{align}\label{eq:Lvo}
 \scr{L}_{\vo} = - \frac{1}{4} \vo^{\mu \nu} \vo_{\mu \nu} + \frac{1}{2} m_{\vo}^2 \vo^\mu \vo_\mu +  \bar{t}\ \gamma_\mu \left[ \goL P_L + \goR P_R \right] T^A V_8^{A, \mu} ~ \! t  \ ,
\end{align}
where $\vo^{\mu \nu} \equiv D_\mu \vo_\nu - D_\nu \vo_\mu $ is the covariant vector-octet field strength tensor with all colour indices omitted for simplicity in both the kinetic and mass terms. As for the vector-singlet case, interactions of the octet state with both the left-handed ($\goL$) and right-handed ($\goR$) top quarks are included, and we consider that both coupling strengths are real. Such simplified models find their motivation, for example, in UV-complete setups featuring compositeness~\cite{Kilic:2009mi,Cacciapaglia:2020kgq} or extra dimensions~\cite{Burdman:2006gy}.

\begin{figure}[t]
\centering
\subfloat[]{%
\includegraphics[width=0.3\textwidth]{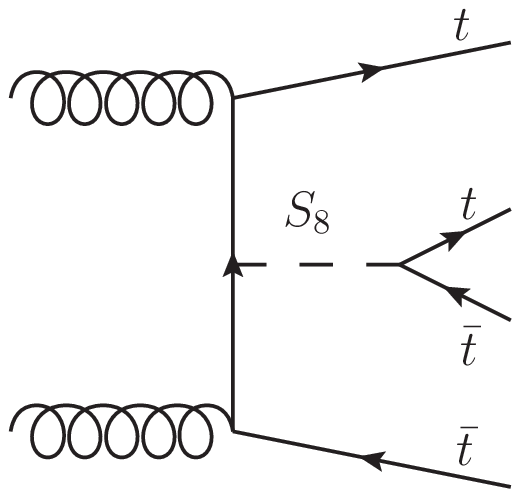}
}%
\hspace{0.2\textwidth}
\subfloat[]{%
\includegraphics[width=0.3\textwidth]{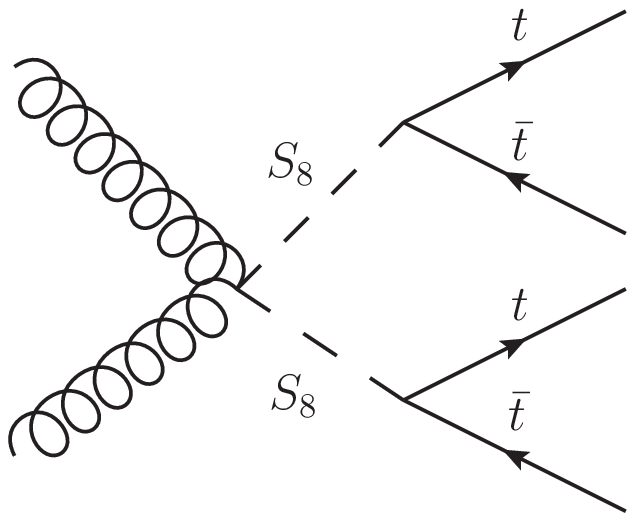}
}%
\caption{Representative Feynman diagrams relevant for four-top production in a simplified model including an extra scalar octet field $\oct$. When the scalar particles are on-shell, the associated production of the scalar octet with a top-antitop pair (a) is proportional to the square of the Yukawa coupling from eq.~\eqref{eq:Lso}, while the QCD-driven pair production of two scalar fields (b) only depends on the octet mass $m_{\oct}$ (Br$(S_8 \to t\bar t) =1$).}
\label{fig:diagram}
\end{figure}

Beyond the different Lorentz structures appearing in the four Lagrangians introduced in this section, the main phenomenological difference between the colour-singlet and colour-octet sets of models is related to the possibility to strongly produce a pair of colour-octet states at hadrons colliders as they directly interact with gluons, already at tree level. This feature is illustrated with the representative Feynman diagrams shown in Fig.~\ref{fig:diagram}. Total cross sections exhibiting large differences in magnitude are thus expected in the different models for a specific new physics mass. This is studied in Sec.~\ref{sec:cs}.

\subsection{Effective higher-dimensional operators in the $SU(3)_{\rm C}\times U(1)_{\rm em}$ basis}\label{sec:EFtmodel}
One of the goals of this work is to compare the modelling of new physics effects that could appear in four-top production through a simplified model approach, such as the ones introduced in Sec.~\ref{sec:model}, with the one originating from the EFT obtained by integrating out the heavy new physics states of the simplified model. We follow the guidelines of ref.~\cite{AguilarSaavedra:2018nen} and focus on the limited set of purely four-top SMEFT operators. In order to ease the matching with the simplified models under consideration, we first re-write the relevant dimension-six SMEFT operators in the Warsaw basis~\cite{Grzadkowski:2010es} in a way invariant under $SU(3)_{\rm C}\times U(1)_{\rm em}$. By introducing the usual notation for the left-handed and right-handed top quark fields, $t_{R/L} \equiv  P_{R/L} t $ with $P_{R/L} = (1 \pm \gamma^5)/2$, we obtain a basis of four independent operators,
\begin{equation}\begin{aligned}\label{eq:OeffEW}
  \oRR= \bar{t}_R \gamma^{\mu} t_R   ~\bar{t}_R \gamma_{\mu} t_R\ , 
  \qquad
  &\oLL= \bar{t}_L \gamma^{\mu} t_L   ~\bar{t}_L \gamma_{\mu} t_L\ ,
  \qquad\\
  \oLR= \bar{t}_L \gamma^{\mu} t_L   ~\bar{t}_R \gamma_{\mu} t_R\ ,
  \qquad
  &\oTaLR= \bar{t}_L T^A  \gamma^{\mu} t_L   ~\bar{t}_R T^A \gamma_{\mu} t_R \ ,
\end{aligned}\end{equation}
where we recall that the $SU(3)$ generators in the fundamental representation are denoted by $T^A$. As we started from a SMEFT formulation that is invariant under electroweak symmetry transformations, so are those operators once we account for the fact that any effective operator involving a left-handed top quark field $t_L$ has to be accompanied by a corresponding operator involving a left-handed bottom quark field $b_L$. The latter operators are however ignored in our study that only examines new physics consequences for four-top production. To define our basis of $SU(2)_{\rm L}$-preserving operators, we have followed the usual conventions, as used for instance in the recent works of refs.~\cite{AguilarSaavedra:2018nen, Hartland:2019bjb, Degrande:2020evl} to which one can relate  using the following mappings\footnote{The main difference with the conventions of  refs.~\cite{Hartland:2019bjb,Degrande:2020evl} stems from  the definition of the $\oLL$ operator  which there includes contributions from SMEFT triplet operators (see ref.~\cite{AguilarSaavedra:2018nen} for more information).}
\begin{align}
    \oRR \leftrightarrow O^1_{tt}\ , \qquad
    \oLR \leftrightarrow O^1_{Qt}\ , \qquad
	\oLL \leftrightarrow \frac12 O^1_{QQ} \ ,\qquad
	\oTaLR \leftrightarrow  O^8_{Qt}\,.
\end{align}
The set of four operators in~\eqref{eq:OeffEW} does not include the SMEFT operator $\oTaLL = \bar{t}_L T^A  \gamma^{\mu} t_L ~\bar{t}_L T^A  \gamma_{\mu} t_L$, which  is redundant when restricted to its pure four-top component, as
\begin{align}
    \oTaLL = \frac{\oLL}{3} \,.
\end{align}
The degeneracy can be lifted  by correlating four-top probes with multi-bottom and multi-top production, as proposed in  ref.~\cite{DHondt:2018cww}. In addition, this relation has been found to approximately hold true also at the next-to-leading order (NLO) in the strong coupling, as bottom quark effects are small~\cite{Degrande:2020evl}.

The matching with the simplified models of Sec.~\ref{sec:model} involves two $SU(2)_{\rm L}$-breaking operators that are by construction not included in the $SU(2)_{\rm L}$-conserving basis~\eqref{eq:OeffEW},
\begin{equation}\label{eq:OeffEWbreaking}
  \os=  \bar{t}  t   ~\bar{t}t\ , \qquad
  \oTas =  \bar{t} T^A  t   ~\bar{t} T^A t  \ .
\end{equation}
The above scalar operators mix left and right-handed top components, and are typically generated from simplified models featuring new scalar particles. As discussed in the previous section, this could arise, for instance, from a 2HDM construction. Such structures cannot be generated by starting from the Warsaw basis of dimension-six SMEFT operators, although they could arise at dimension eight, for instance via a ``Yukawa squared'' operator $(H \cdot Q_L t_R) \, (H \cdot Q_L t_R)$.

It is worth mentioning that a complete basis sufficient to describe {\it any} operator invariant under the $SU(3)_{\rm C}\times U(1)_{\rm em}$ symmetry should also include the two parity-breaking operators \begin{equation}\begin{aligned}
  \ops=  \bar{t}  t   ~\bar{t} i \gamma^5 t\ , \qquad
  \oTaps =  \bar{t} T^A  t   ~\bar{t} T^A  i \gamma^5  t  \ .
\end{aligned}\end{equation}
These two operators are, however, irrelevant in the context of the matching with the simplified models listed above, and will thus be ignored in the following.\footnote{Such terms arise in models featuring scalar fields interacting with both a  scalar and pseudo-scalar top current. They can be generated, for instance, in a simplified model involving a new complex scalar field with complex Yukawa couplings.} In conclusion, as far as four-top final states are considered (and possible contributions from $b$ quarks in the initial state are neglected) the basis introduced above is complete and therefore suitable to generically study the dynamics of four-top quark interactions. 
 
Any four-top operator invariant under the $SU(3)_{\rm C}\times U(1)_{\rm em}$ symmetry can thus be related to the basis of operators,
\begin{equation}
    \Big\{ \oRR, \oLL, \oLR, \oTaLR, \os, \oTas, \ops, \oTaps \Big\}\ .
\label{eq:fullbasis}\end{equation}
The first ingredient behind this demonstration is the Fierz decomposition of a fermion bispinor $\fa \fab$, where $i$ and $j$ are colour indices,
\begin{align}
    \fa \fab = - \frac{1}{4} \left( \fab \fa\ \mathbb{1} +  \fab  \gamma^5 \fa  \ \gamma^5 + \fab \gamma^\mu \fa \ \gamma_\mu - \fab \gamma^\mu\gamma^5 \fa \ \gamma_\mu \gamma^5  + \frac{1}{2} \fab \sigma^{\mu \nu} \fa\ \sigma_{\mu \nu} \right) \ ,
\end{align}
in which we use the standard definition $\sigma^{\mu \nu} = \frac{i}{2} [ \gamma^\mu,  \gamma^\nu]$. As we deal with coloured objects, we further need to rely on  $SU(3)$ algebra relations allowing us to swap the colour indices,
\begin{align}
     T^A_{kj} T^A_{il} = \frac{4}{9} \delta_{ij} \delta_{kl} - \frac{1}{3} T^A_{ij} T^A_{kl}\ , \qquad
      \delta_{kj} \delta_{il} = \frac{1}{3} \delta_{ij} \delta_{kl} +2 T^A_{ij} T^A_{kl} \ .
\end{align}
We refer the interested reader to App.~\ref{sec:matching} for details.

\begin{table}\begin{center}
  \setlength\tabcolsep{8pt}
  \renewcommand{\arraystretch}{2.0}
  \begin{tabular}{ c c | c c c c c c}
   \multicolumn{2}{c|}{Heavy Mediator}	 & $\os$ & $\oTas$ & $\oLL$  &  $\oRR$ &  $\oLR$ & $\oTaLR$\\
   \hline
   \multirow{3}{*}{\rule{0pt}{1.95em} Colour Singlets}  &$\sing$ & $\ds\frac{\ysS^2}{2 m_{\sing}^2}$  & / &/ &  / & / & / \\
   &		$\tilde{\sing}$ & $ \ds  - \frac{\ysP^2}{2  m_{\tilde{\sing}}^2}$  & / &/ &  / &  $ \ds  - \frac{\ysP^2}{3 m_{\tilde{\sing}}^2}$ &  $\ds   - 2 \frac{\ysP^2}{m_{\tilde{\sing}}^2}$ \\
   &	$\vs$ & / & / & $\ds  -\frac{\gsL ^2}{2 m_{\vs}^2}$& $\ds  -\frac{\gsR^2}{2 m_{\vs}^2} $ &  $\ds   - \frac{\gsL \gsR}{m_{\vs}^2}$ & / \\[.75em]
   \hline
   \multirow{3}{*}{\rule{0pt}{1.95em} Colour Octets}   &$\oct$ & / & $\ds \frac{\yoS^2}{2 m_{\oct}^2}$ & / & / & / & / \\
	&	$\tilde{\oct}$ & / &$\ds  - \frac{\yoP^2}{2 m_{\tilde{\oct}}^2}$  & / &  / &  $\ds  - \frac{4 \yoP^2}{9 m_{\tilde{\oct}}^2}$ &  $\ds  \frac{\yoP^2 }{3 m_{\tilde{\oct}}^2} $\\
	&	$\vo$ & / & / & $\ds  -\frac{\gsL^2}{6 m_{\vo}^2}  $& $\ds  -\frac{\gsR^2}{6 m_{\vo}^2} $ &  / & $\ds -\frac{\goL \goR}{ m_{\vo}^2} $
  \end{tabular}
  \caption{Wilson coefficients obtained when the various simplified top-philic models described in Sec.~\ref{sec:model} are matched with the six $P$-conserving operators of the EFT basis~\eqref{eq:fullbasis}. The entries in this table represent all Wilson coefficients $c_i$ corresponding to an EFT Lagrangian ${\cal L}_{\rm EFT}({\tphi}) = \sum\limits_{i} c_i({\tphi})\ \mathcal{O}_i$ once the heavy top-philic mediator $\tphi$ is integrated out.} 
\label{tab:match}\end{center}\end{table}

In order to relate the simplified models of Sec.~\ref{sec:model} to the EFT basis~\eqref{eq:fullbasis}, we assume that all introduced coupling strengths are real, as already mentioned, and we neglect any renormalisation group running of the operator coefficients. The latter is in particular justified as we enforce the matching of the simplified model with the EFT at the mass scale of the new states, so that only small scale separations are in order (taming thus any potential renormalisation group effect). We leave the details on the matching procedure to App.~\ref{sec:matching}, and present our results in Table~\ref{tab:match}, using the notations of the previous section for the couplings. This table provides an expression of the Wilson coefficients $c_i$ appearing in the EFT Lagrangian ${\cal L}_{\rm EFT}({\tphi})$, with 
\begin{align}
{\cal L}_{\rm EFT}({\tphi}) = \sum\limits_{i} c_i({\tphi})\ \mathcal{O}_i\ ,
\end{align} as a function of the parameters of the various simplified models once the heavy top-philic mediator $\tphi$ is integrated out. In our results, we consider the limiting cases of new scalar states that are either $CP$-even (field names without a tilde) or $CP$-odd (field names with a tilde). One emerging feature concerns the sign of the resulting effective Wilson coefficients, that differs between scalar and vector models. This can be traced to the opposite sign of the $\tphi$ propagator appearing in the diagrams computed in the UV theory when the large new physics scale limit is taken.

\subsection{Four-top production cross sections}
\label{sec:cs}
\subsubsection{Generalities}
In the considered top-philic simplified models, four-top production at the LHC involves three main channels: i) the QCD-driven production of a pair of  top-philic states ($pp\to \tphi \tphi$) followed by their decay into a $t\bar t$ system, relevant only when the $X$ state is a colour-octet, ii) the associated production of a single top-philic particle with a $t\bar t$ pair ($pp\to t\bar t \tphi$) followed by an $\tphi\to t\bar t$ decay, and iii) off-shell four-top production (without any {\it on-shell} intermediate $\tphi$ resonance). We illustrate these processes in the case of a scalar-octet simplified model with the Feynman diagrams shown in Fig.~\ref{fig:diagram}. In order to properly include the interference of these new physics contributions with the SM contributions entering at different ${\cal O}(\alpha^k \alpha_S^n)$, with $\alpha_S$ and $\alpha$ denoting the strong and weak coupling respectively, we make use of the \amc\ platform~\cite{Alwall:2014hca}. We estimate the new physics effects by first implementing the models introduced above into \fr, and by then converting them into the UFO format~\cite{Christensen:2009jx,Alloul:2013bka,Degrande:2011ua}.

We begin with the evaluation of the four-top production cross section at the LHC, at a centre-of-mass energy $\sqrt{S}=13$~TeV, for all new physics scenarios considered here.  In order to assess the theoretical uncertainties inherent to our calculations, we set the factorisation and renormalisation scales to a common value $\mu$ that varies in the range
\begin{align}
\label{eq:muscale}
    \mu \in [\sqrt{\hat s}/4, \sqrt{\hat s}] \ ,
\end{align}
where $\sqrt{\hat s}$ is the partonic centre-of-mass energy. The central value $\mu_0 = \sqrt{\hat s}/2$ is chosen to allow for a smooth comparison of the simplified model predictions (that involve on-shell heavy particle production) with the EFT predictions that are expected to be only reliable at high energies. Such a scale is significantly larger than the one chosen in ref.~\cite{Degrande:2020evl}, in which $\mu_0\sim 2 m_t$, as the latter work solely focused on a comparison with $t\bar t$ production (for which the top mass is the natural scale). All the results presented in this work rely on leading-order (LO) cross sections, and our predictions include QCD and electroweak (EW) SM contributions, new physics contributions, as well as their interference (see App.~\ref{sec:MC} for some technical details). This cross section will be called Complete LO (C-LO) in the rest of this work. We stress that both LO and C-LO predictions are obtained at ``tree-level'', and do not include NLO corrections. The only difference between both choices is the inclusion or not of the interference with the EW diagrams. Complete next-to-leading-order (C-NLO) calculations ({\it i.e.}~including all EW and QCD corrections) are available for SM four-top production, yet are currently out of reach for new physics scenarios. 

In addition, while next-to-leading-order (NLO) QCD corrections are technically feasible, they do not suffice to give a reliable estimate of the cross sections, as electroweak contributions (already at tree-level) turn out to be important. As an approximation, one could consider using the C-LO results in the new physics scenarios and derive an NLO estimate of the complete four-top cross section $\sigma^{\rm NP}_{4t}$ (in the sense that all QCD, electroweak and new physics diagrams are included). The SM C-NLO predictions are indeed known, with $\sigma_{\rm SM}^{\rm{C-NLO}} = 11.97^{+2.15}_{-2.51}~{\rm fb}$~\cite{Frederix:2017wme}, such that a ``multiplicative'' scheme could be implemented. In such a scheme, the approximated C-NLO new physics cross section $\tilde\sigma$ would be given by
\begin{align}\label{eq:multscheme}
  \tilde\sigma^{\rm C-NLO}_{\rm NP} \approx \sigma_{\rm SM}^{\rm{C-NLO}} \times \left( \frac{ \sigma_{\np}^{\rm{LO}} }{ \sigma_{\rm SM}^{\rm{LO}} } \right)\equiv K_{\rm SM} \, \sigma_{\np}^{\rm{LO}}  \ ,
\end{align}
where the multiplicative factor is taken as the ratio of the LO pure new physics contributions ($\sigma_{\np}^{\rm{LO}}$) to the SM ones ($\sigma_{\rm SM}^{\rm{LO}}$). However, simple observations suggest that such an estimate could be unreliable, and therefore  we refrain from adopting it. The SM $K$-factor $K_{\rm SM}$ ({\it i.e.}~the ratio of the C-NLO SM cross section to the C-LO one) is 2.3, whilst pure QCD $K$-factors associated with the EFT contributions range from 0.9 to 1.4~\cite{Degrande:2020evl}. EW $K$-factors for the EFT cross sections are presently unknown. Concerning the simplified models examined in our study, a QCD $K$-factor close to $2$ was numerically obtained for the $pp\to \oct \oct$ process~\cite{Degrande:2014sta,Darme:2018dvz}, while there is no similar result for the rest of the new physics scenarios considered here. Given our present knowledge, using $K_{\rm SM}$ seems just a (possibly wild) guess. For this reason, we prefer to explicitly acknowledge our ignorance and use C-LO cross sections as reference values. Moreover, we will vary the normalisation by adding a factor between $1$ and $2$ to account for unknown higher-order uncertainties. 

We further refer to App.~\ref{sec:EWint} for more details regarding the importance of including the interference with SM electroweak contributions and close this section by presenting a semi-analytical approximation of the EFT cross sections,
\begin{align}
\label{eq:approxval}
 \sigma^{4t \rm EFT}_{\rm NP} \ (\rm  in  \ \rm fb) =~\begin{cases}
\displaystyle \ 0.57  ~\rm{TeV}^4~\frac{y_{1S}^4}{\Lambda^4}+0.064~\rm{TeV}^2 ~\frac{ y_{1S}^2}{\Lambda^2} \quad \textrm{  (scalar singlet)} \\[0.6em]
\displaystyle \ 2.36 ~\rm{TeV}^4~\frac{g^4}{\Lambda^4}+ 0.47 ~\rm{TeV}^2~\frac{g^2}{\Lambda^2} \qquad \textrm{ (vector singlet)} \\[0.6em]
\displaystyle \ 0.13~\rm{TeV}^4~\frac{y_{8S}^4}{\Lambda^4} - 0.15 ~\rm{TeV}^2~\frac{y_{8S}^2}{\Lambda^2} \quad\ \textrm{ (scalar octet)} \\[0.6em]
\displaystyle \ 0.55 ~\rm{TeV}^4 ~\frac{g^4}{\Lambda^4}+0.63 ~\rm{TeV}^2 ~\frac{g^2}{\Lambda^2} \qquad \textrm{ (vector octet)} 
 \end{cases} \, ,
\end{align}
where we have introduced $g_{1L}=g_{1R} = g_{8L}=g_{8R} \equiv g$ in the matching conditions of Table~\ref{tab:match}. For couplings of ${\cal O}(1)$, the interference term starts to dominate for $\Lambda\sim 1$~TeV for most models, with the exception of the scalar singlet case.

\subsubsection{Predictions for the four-top total cross sections}
\label{sec:4tCS}
\begin{figure}[t]
\centering
\subfloat[]{%
\includegraphics[width=0.49\textwidth]{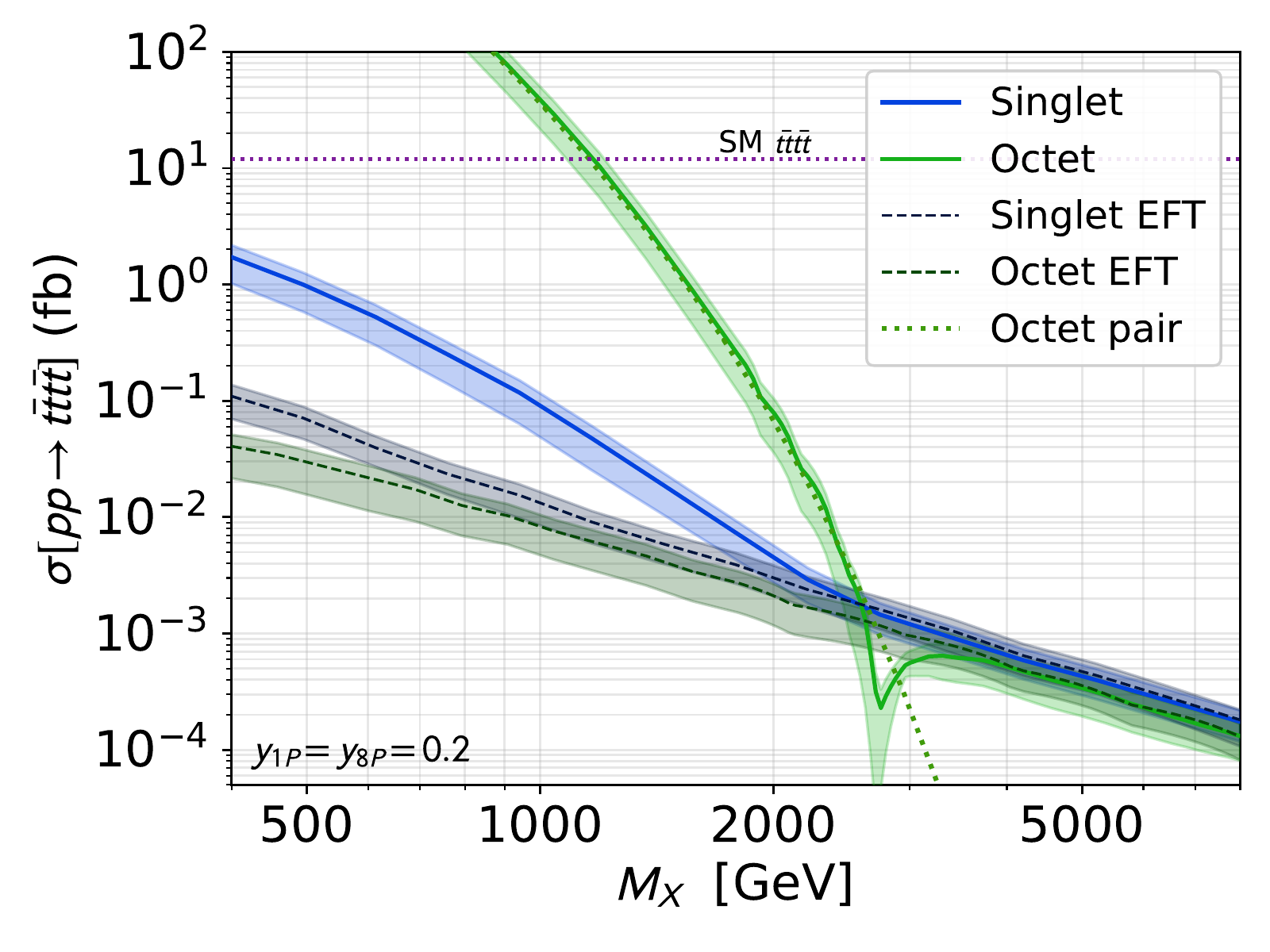}
}%
\subfloat[]{%
\includegraphics[width=0.49\textwidth]{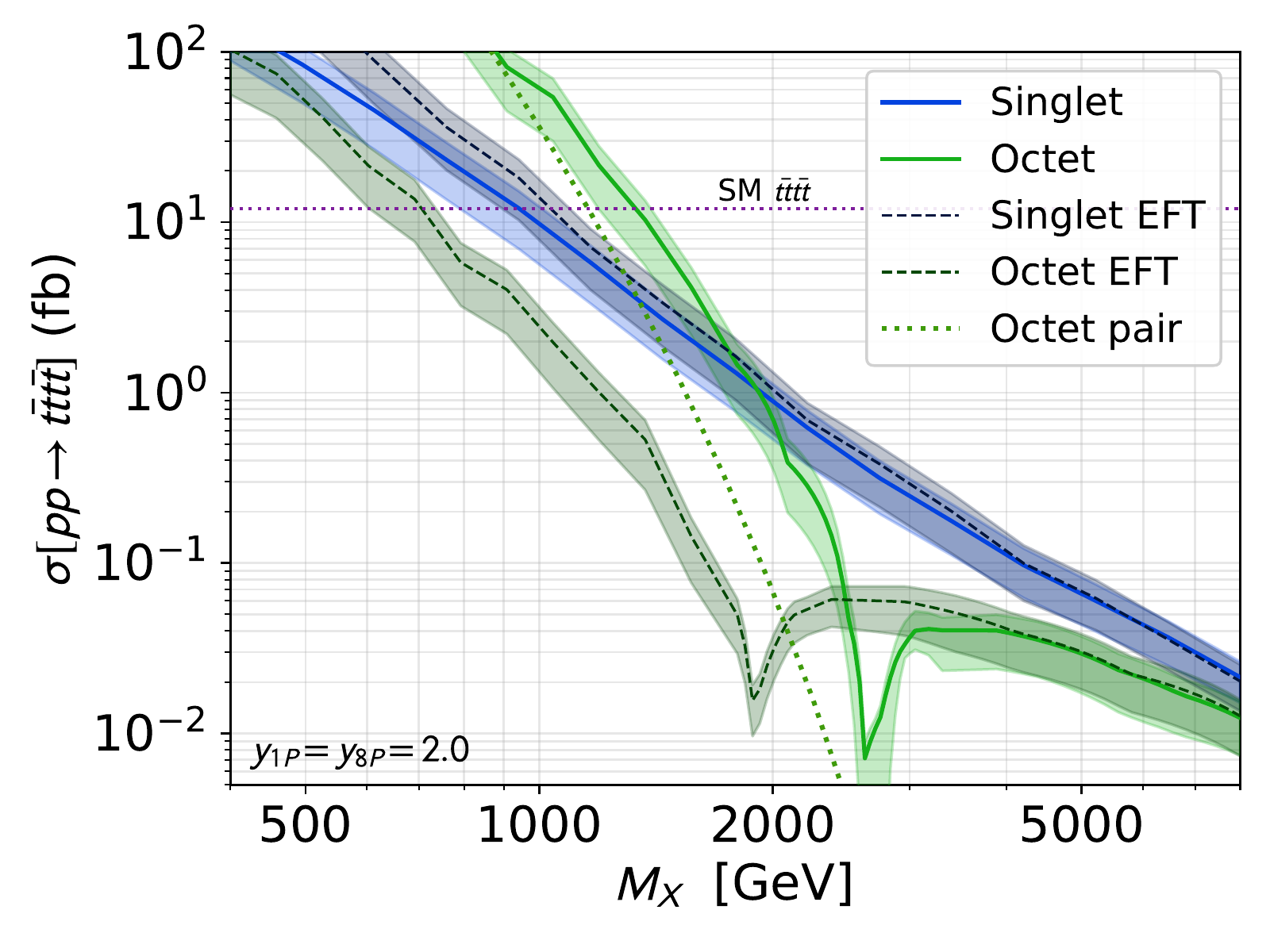}
}%
\hspace{0.01\textwidth}
\subfloat[]{%
\includegraphics[width=0.49\textwidth]{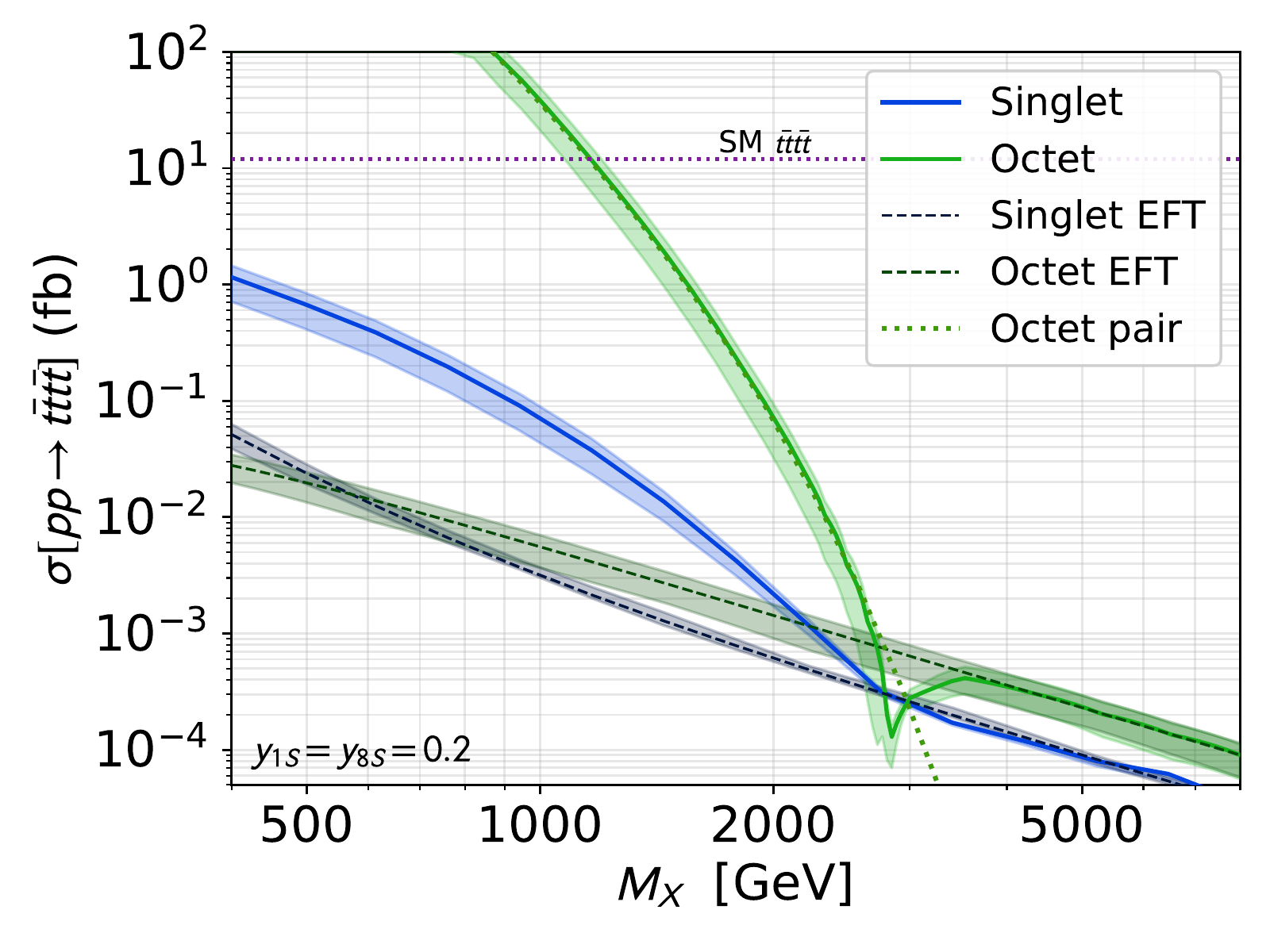}
}%
\subfloat[]{%
\includegraphics[width=0.49\textwidth]{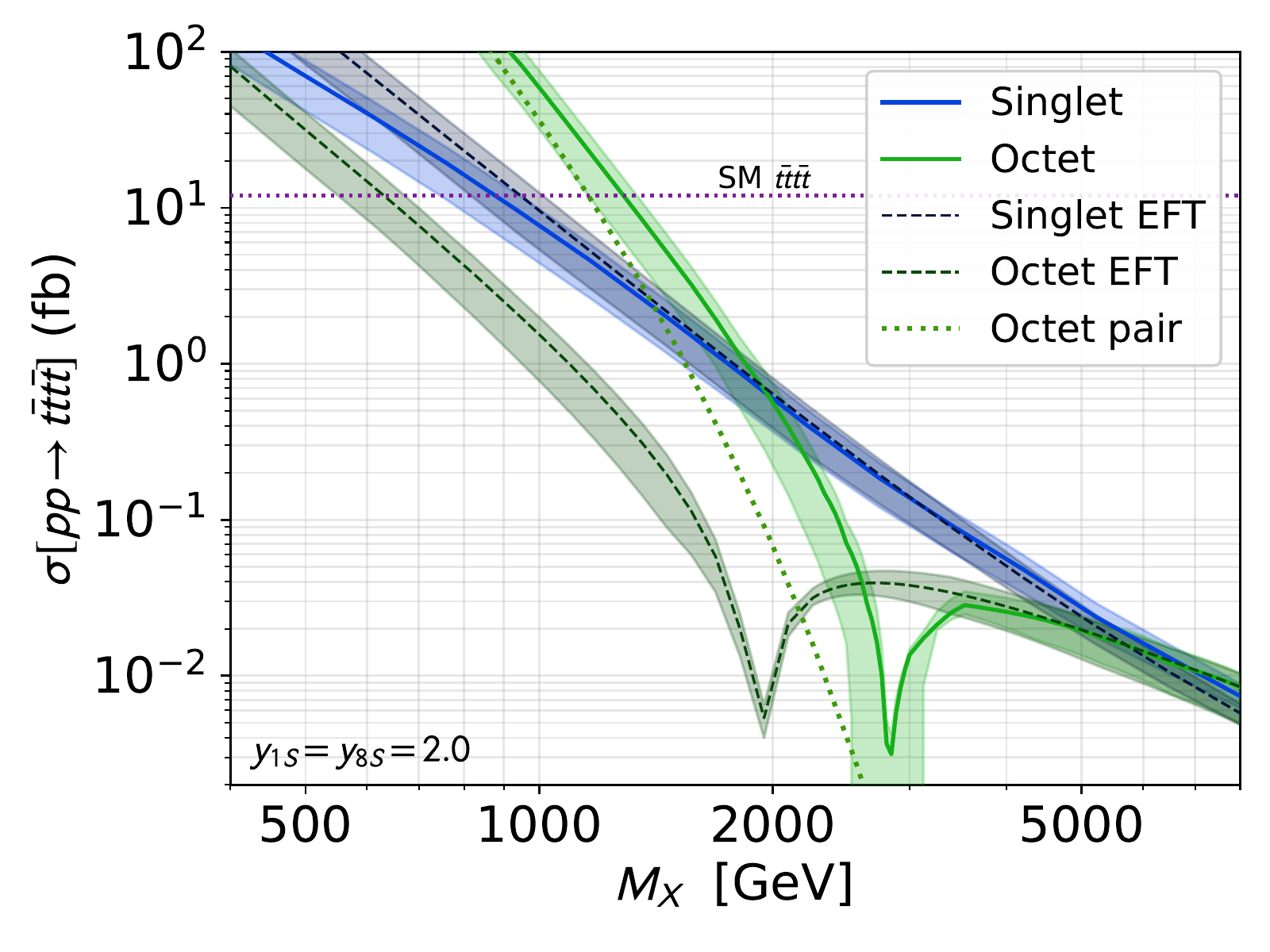}
}%
\caption{New physics contributions (including their interference with the SM component) to the four-top production cross section as function of the top-philic particle mass $M_{\tphi}$ in the simplified model (thick lines) and EFT (dashed lines) approaches, for the case of a top-philic colour-singlet (blue) and octet (green) scalar field, and for different choices of the new physics couplings defined in eqs.~\eqref{eq:Lss} and \eqref{eq:Lso}. We consider both a pseudo-scalar case (a, b) and a scalar case (c, d), and coupling values of 0.2 (a, c) and 2 (b, d). The theoretical uncertainties are obtained as described by eq.~\eqref{eq:muscale}, and the green dotted line represents the $p p \to \oct \oct$ production cross section. We indicate by a purple dotted line the SM $t \bar t t \bar t$ cross section from ref.~\cite{Frederix:2017wme}.}
\label{fig:CSscalarInt}
\end{figure}
We now turn to the study of the predictions for the four-top production cross section in the models studied in this paper. In Fig.~\ref{fig:CSscalarInt}, we show the dependence of the C-LO four-top production cross section on the top-philic scalar (upper row) and pseudo-scalar (lower row) mass for two different coupling values of 0.2 (left figures) and 2 (right figures). The theoretical uncertainties associated with each curve are obtained by scale variations as described by eq.~\eqref{eq:muscale}. 

\begin{figure}[t]
\centering
\subfloat[]{%
\includegraphics[width=0.47\textwidth]{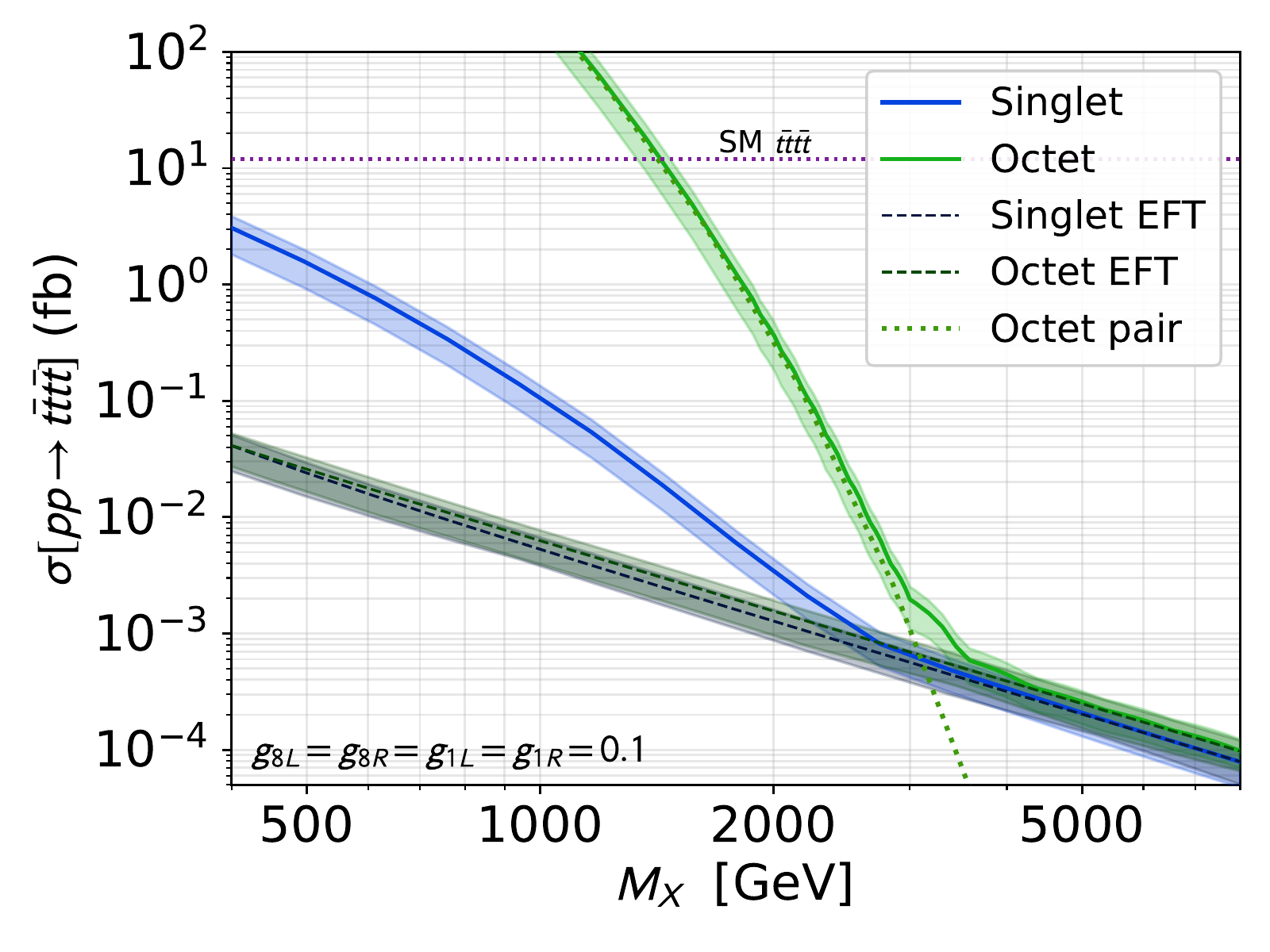}
\hspace{0.02\textwidth}
\subfloat[]{%
\includegraphics[width=0.47\textwidth]{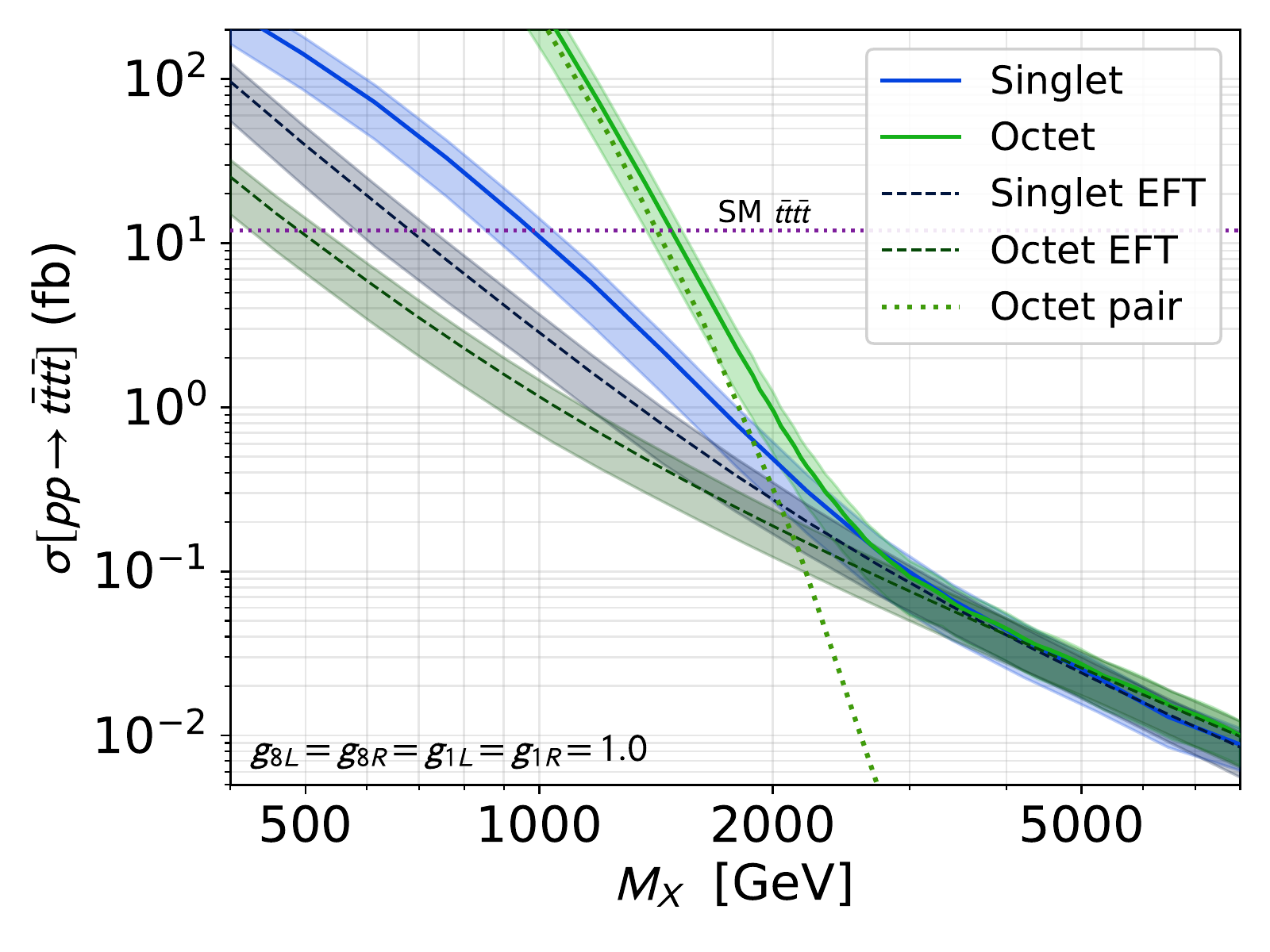}
}%
}%
\caption{Four-top production cross section as function of the top-philic particle mass $M_{\tphi}$ in the simplified model (thick lines) and EFT (dashed lines) approaches, for the case of a vector top-philic singlet (blue) and octet (green) field, and for different choices of the new physics couplings defined in eqs.~\eqref{eq:Lvs} and \eqref{eq:Lvo}. We consider the cases of $\gsL = \gsR = \goL = \goR = 0.1$ (a) and $\gsL = \gsR = \goL = \goR = 1$ (b). The theoretical uncertainties are obtained as described by eq.~\eqref{eq:muscale}, and the green dotted line represents the $p p \to \vo \vo$ production cross section. We indicate by a purple dotted line is the SM $t \bar t t \bar t$ cross section from ref.~\cite{Frederix:2017wme}.}
\label{fig:CSvectorInt}
\end{figure}

On-shell production of the new top-philic fields, either via pair-production or through their associated production with a $t\bar t$ pair, dominates at low masses and small couplings. This effect is particularly visible in Figs.~\ref{fig:CSscalarInt} (a) and (c), in which we can observe that the simplified model predictions converge to the EFT ones only once the top-philic particles $\tphi$ become too heavy to be resonantly produced. As a further illustration, we show as a green dotted line the cross section associated with scalar-octet pair production. This cross section steeply falls with the mass, as typical for any QCD-induced production of a pair of heavy coloured states. As a consequence, the large negative interference between the new physics diagrams and the SM diagrams yield the visible V-shape behaviour at $M_X\sim 2.5$~TeV that is not present for singlet top-philic states. The new physics contributions to the cross section are positive for smaller masses and become negative for masses above about 2.5~TeV. Such a feature is expected from eq.~\eqref{eq:approxval}. It however occurs for cross section values already well below the current LHC and future HL-LHC sensitivities, so that it will not have any impact in practice. The situation is relatively similar when dealing with vector fields, as shown in Fig.~\ref{fig:CSvectorInt}. All these figures additionally validate numerically the matching relations of Table~\ref{tab:match}, as they demonstrate that the simplified model and EFT cross sections agree with each other for very heavy top-philic states $\tphi$. 

We stress that the dominance of the double and single  resonant production at low masses and weak couplings is also a consequence of the minimal structure of the simplified model, which implies  BR($\tphi \to t \bar t)=1$. With other decay channels being accessible to the $\tphi$ state, the relative importance of resonant production could be drastically suppressed. For instance,  a top-philic scalar singlet that is the mediator to a dark sector, coupling to a dark matter candidate $\chi$, can have a large branching ratio BR($\tphi \to \chi \bar \chi)$, as shown for example in ref.~\cite{Arina:2016cqj}.

When examining benchmark scenarios for which the coupling is large, we need to keep in mind that the width of the top-philic particle $\tphi$ also increases. This effect is especially relevant for the scalar singlet field, which has $\Gamma_{\sing} \sim M_{\sing}/2$ for Yukawa couplings around $2$ within the model definition~\eqref{eq:Lss}. For $\Gamma_X/M_X= {\cal O}(10 \%)$, the production and decay of the top-philic states interfere with the SM background and cannot be factorised as in the narrow-width approximation. For larger widths, the very approach of using a fixed width might even not be adequate.

\begin{figure}[t]
\centering
\subfloat[]{%
\includegraphics[width=0.6\textwidth]{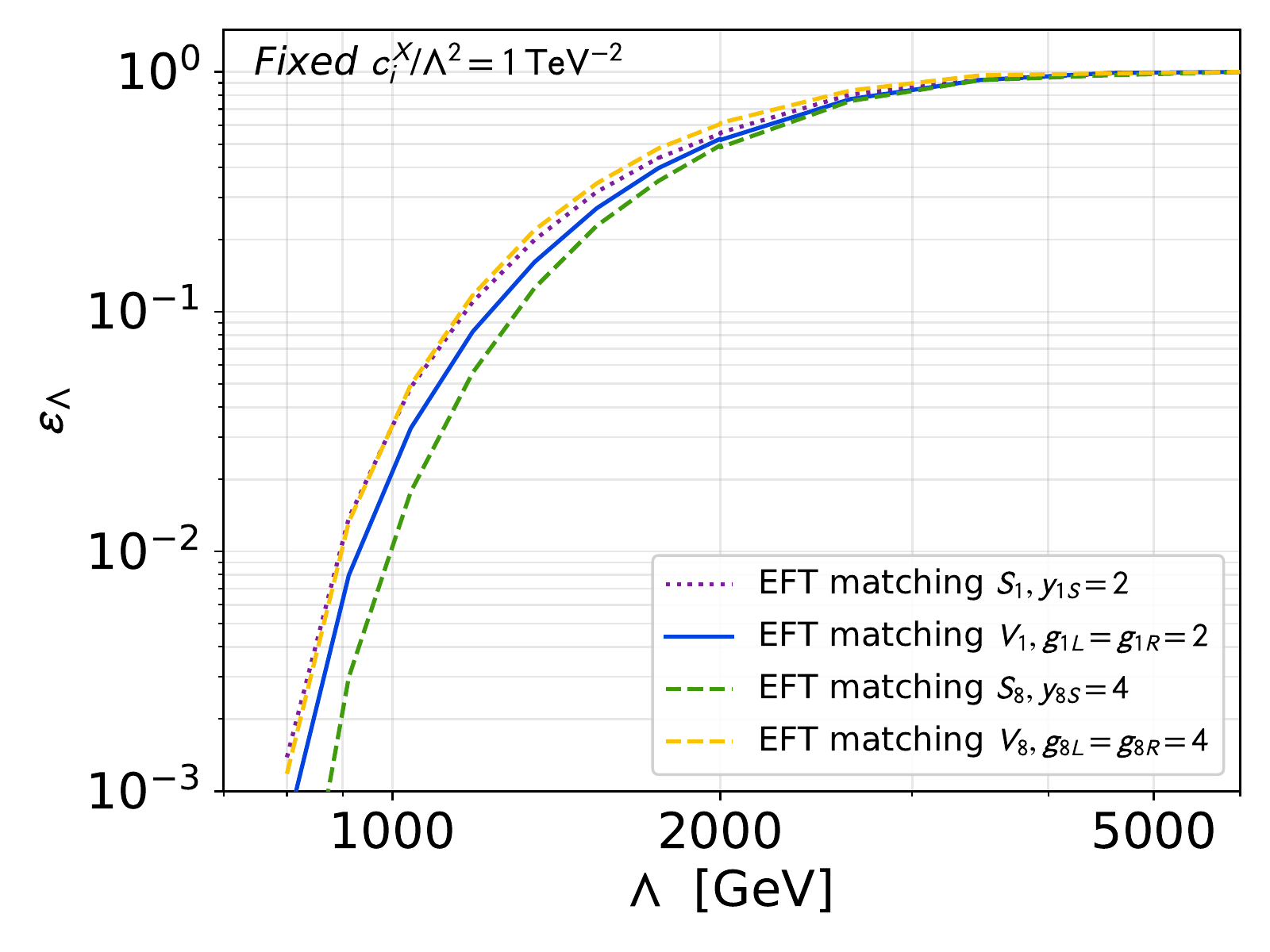}
}%
\caption{Efficiency~\eqref{eq:lambdaeff} obtained by varying the cut-off scale at which the high-energy behaviour of the cross section in the EFT is truncated. The non-truncated rate $\sigma^{\rm EFT}_{4t} (c_i({\tphi}), \infty)$ is evaluated for $c_i(\tphi) / \Lambda^2 = 1$~TeV$^{-2}$. We present results for various choices of simplified models.
\label{fig:EffEFTScale}}
\end{figure}

When estimating the EFT cross section at low scales, the EFT scale $\Lambda$  may become comparable with the partonic centre-of-mass energy. In this case, extra care should be taken in the interpretation, as on-shell production might dominate in such a regime. This is illustrated in  Figs.~\ref{fig:CSscalarInt} and \ref{fig:CSvectorInt} by the green dashed lines. Following the current recommendations (see, \eg, refs.~\cite{Racco:2015dxa, AguilarSaavedra:2018nen}),  a possible procedure to make meaningful interpretations of  EFT predictions  is to implement an event-per-event cut on the partonic centre-of-mass energy $\sqrt{\hat s}$, ensuring that it remains below the EFT scale. In order to assess the impact of such a procedure, we make use of the expressions of the EFT coefficients $c_i({\tphi})$ in terms of the simplified model parameters introduced in Table~\ref{tab:match}. For example, a scenario including a vector singlet state with couplings set to $\gsL = \gsR =1$ leads to the six Wilson coefficients values
\begin{equation}
    c_i({V_1}) =  \Big(c^1_S(V_1), c^8_S(V_1), c^1_{LL}(V_1), c^1_{RR}(V_1), c^1_{LR}(V_1), c^8_{LR}(V_1)\Big) = \Big(0, 0, -0.5, -0.5, -1, 0\Big)\, ,
\end{equation}
once the heavy top-philic state is integrated out. Making use of the resulting EFT, we estimate the impact of the truncation of the high-energy behaviour of the cross section through cross section computations while enforcing a dynamical cut $\sqrt{\hat s} < \Lambda$. In practice, we first evaluate the EFT four-top total rate $\sigma^{\rm EFT}_{4t}(c_i({\tphi}), \infty)$ without the implementation of any cut on the high-energy behaviour of the cross section. Then, we re-calculate the rate by varying $\Lambda$ together with the Wilson coefficients so that the ratios $c_i({\tphi}) / \Lambda^2$ are kept constant (in doing so, the cross sections $\sigma^{\rm EFT}_{4t}(c_i({\tphi}), \infty)$ do not change). We next compute the rates $\sigma^{\rm EFT}_{4t}(c_i({\tphi}), \Lambda)$ including the cut $\sqrt{\hat s} < \Lambda$, and finally evaluate the efficiencies $\epsilon_\Lambda$ defined by
\begin{align}
    \epsilon_\Lambda ~\equiv~    \frac{\sigma^{\rm EFT}_{4t} (c_i({\tphi}), \Lambda)}{\sigma^{\rm EFT}_{4t} (c_i({\tphi}), \infty)} \,.
\label{eq:lambdaeff}\end{align}
The dependence of these efficiencies on $\Lambda$ is presented in Fig.~\ref{fig:EffEFTScale} for various choices of simplified models and couplings. It is found that they only exhibit a small dependence on the choices of effective operators (and thus on the simplified model parameters). This reflects the behaviour of the parton distribution functions at large $x$ that turns out to be more important than details of the new physics model.

In practice, Fig.~\ref{fig:EffEFTScale} could be used to derive conservative EFT limits from four-top measurements. However, we will not implement such a scheme in our work, firstly because our goal is precisely to compare \textit{naive} EFT estimates with the simplified model approach. Secondly, as will be shown below, the naive EFT approach has the tendency to underestimate any limit on a new physics theory from a direct comparison with LHC results. In any case it is clear from the results presented above that the four-top EFT approach does not accurately reproduce the behaviour of the simplified models considered in this work in the parameter space accessible at the LHC. However, at least in some cases, the EFT treatment might still provide a useful estimate of the typical experimental efficiencies associated with a four-top search. When this holds, it opens the door to a variety of reinterpretations of the results of a single EFT analysis by simple cross section rescalings.


\section{Recasting experimental four-top searches}\label{sec:recast}
In order to assess the impact of the most recent LHC results in the four-top-quark channel on new physics, we use the implementation~\cite{Darme:2020hxc,Fuks:2021wpe} in the \ma~framework~\cite{Conte:2012fm,Conte:2014zja,Dumont:2014tja,Conte:2018vmg} of the CMS-TOP-18-003 analysis~\cite{Sirunyan:2019wxt}. This analysis investigates four-top production in final states including multiple leptons, light jets and $b$-tagged jets with a luminosity of 137~fb$^{-1}$ of $pp$ collisions at a centre-of-mass energy of $\sqrt{S} = 13$~TeV. 

We make use of our \fr/UFO implementations~\cite{Christensen:2009jx,Degrande:2011ua,Alloul:2013bka} of the simplified models of Sec.~\ref{sec:model} and the EFT operators of Sec.~\ref{sec:EFtmodel} to generate hadron-level events for each model within the \amc\ platform~\cite{Alwall:2014hca}. Hard-scattering event generation relies on the convolution of LO matrix elements with the NNPDF3.0 set of parton densities~\cite{Ball:2014uwa,Buckley:2014ana}. The resulting events are matched with parton showers as handled by \py~\cite{Sjostrand:2014zea}, which we also use for the simulation of hadronisation. The simulation of the response of the CMS detector and event reconstruction are performed with \del~\cite{deFavereau:2013fsa} (with an appropriate detector parametrisation), that internally relies on \fj~\cite{Cacciari:2011ma} and its implementation of the anti-$k_T$ algorithm~\cite{Cacciari:2008gp}. In our simulation chain, the whole detector simulation and event reconstruction process is driven by \ma~\cite{Conte:2012fm,Conte:2014zja,Dumont:2014tja,Conte:2018vmg}, which is then used for extracting the efficiencies relevant for the various signal regions of the analysis.

We briefly describe in Secs.~\ref{sec:macms} and \ref{sec:highHT} the CMS-TOP-18-003 analysis and its implementation and validation in \ma, and how we design from this Standard Model analysis a signal region targeting specifically new physics. Details on our numerical limit extraction procedure are given in Sec.~\ref{sec:fullnumerics}.

\subsection{The CMS-TOP-18-003 search and its implementation in \ma}\label{sec:macms}
As a detailed description of the validation of the implementation in \ma\ of the CMS-TOP-18-003 analysis has been published separately~\cite{Darme:2020hxc,Fuks:2021wpe}, we only briefly review in this section its most relevant ingredients.

The CMS-TOP-18-003 analysis targets specific final states that can originate from four-top production and decay. Once produced, the four top quarks all decay into a $b$-quark and a $W$-boson, the latter decaying then either leptonically or hadronically. Four-top production hence leads to final states featuring between zero and four leptons at the hard-scattering level. The CMS-TOP-18-003 search focuses on final states featuring either a pair of leptons (electrons or muons) with the same electric charge, or three or more leptons. Both the requirement of a same-sign lepton pair and the one of a large number of leptons allow for a significant reduction of the SM top-antitop background, as enforcing the selected events to include a large multiplicity of $b$-tagged and non-$b$-tagged jets.

\begin{table}\begin{center}
  \renewcommand{\arraystretch}{1.25}
  \begin{tabular}{@{}cccccc@{}} 
    \multicolumn{6}{l}{\rule{0pt}{1.25em}  \textbf{Basic kinematic requirements}} \\
    \hline  & Electrons & Muons & Jets & $b$-tagged jets &\\
      $p_T$ (GeV) & $>20$ &  $>20$ &  $>40$ &  $>25$ &\\
      $|\eta|$ & $<2.5$ &  $<2.4$ &  $<2.4$ &  $<2.4$&\\[0.5em]
   \multicolumn{6}{l}{ \rule{0pt}{1.25em}  \textbf{Baseline selection}} \\
   \hline Jets & \multicolumn{5}{p{13cm}}{$H_T > 300$ GeV,  $p_T^{\text{miss}} > 50$ GeV, at least two  jets and two $b$-tagged jets} \\
    Leptons& \multicolumn{5}{p{13cm}}{If same charge pair: $p_T(\ell_1) > 25$ GeV and $p_T(\ell_i) > 20$ GeV for $i\neq 1$} \\
Isolation & \multicolumn{5}{p{13cm}}{Jets and $b$-tagged jets: $\Delta R > 0.4$ with respect to the selected leptons} \\[0.5em]
    \multicolumn{6}{l}{\textbf{\rule{0pt}{1.25em}  Further vetoes}} \\
   \hline Vetoed & \multicolumn{5}{p{13cm}}{Same-sign electron pairs with an invariant mass smaller than $12$ GeV} \\
    Vetoed & \multicolumn{5}{p{13cm}}{Third lepton with $ p_T > 5\  (7)$ GeV for $e$ ($\mu$) forming an opposite-sign same-flavour pair with an invariant mass satisfying $m_{\text{OS}} < 12$ GeV or $m_{\text{OS}} \in [76,106]$~GeV}
  \end{tabular}
    \caption{Summary of the selection cuts of the CMS-TOP-18-003
    analysis~\cite{Sirunyan:2019wxt}. We define $H_T$ as the sum of the transverse
    momenta of all jets.}\label{tab:preselection} \vspace{.7cm}

  \setlength\tabcolsep{8pt}
  \begin{tabular}{@{}cccc|c c @{}} 
  \rule{0pt}{1.25em}
     $N_\ell$  & $N_b$ & $N_j$ & Region & $t\bar{t} t\bar{t}$ (SM signal - CMS)  & $t\bar{t} t\bar{t}$ (Background - CMS)  \\
     \hline   \rule{0pt}{1.25em}2 & 3       & 6        & SR5 & $1.61\pm 0.90 $ & $5.03\pm 0.77$\\
     2 & 3       & 7        & SR6 & $1.14\pm 0.66 $  & $2.29\pm 0.40$\\
     2 & 3       & $\geq$8  & SR7 & $0.85\pm 0.47 $  & $0.71\pm 0.20$\\
     2 & $\geq4$ & $\geq$5  & SR8 & $2.08\pm 1.23 $  & $3.31\pm 0.95$\\[.2cm]
     $\geq 3$ & $\geq 3$ & 4        & SR12 & $0.56\pm 0.32 $  & $2.03\pm 0.48$\\
     $\geq 3$ & $\geq 3$ & 5        & SR13& $0.66\pm 0.38 $   & $1.09\pm 0.28$\\
     $\geq 3$ & $\geq 3$ & $\geq$6  & SR14 & $0.76\pm 0.45 $  & $0.87\pm 0.30$
  \end{tabular}
  \caption{Definition of the most relevant signal regions of the CMS-TOP-18-003 analysis, together with the expectation from the SM $t\bar{t} t\bar{t}$ signal and background, as reported by the CMS collaboration (pre-fit results are used)~\cite{Sirunyan:2019wxt}.
    }\label{tab:sr} 
\end{center}\end{table}

We summarise in Table~\ref{tab:preselection} the selection criteria used in the CMS-TOP-18-003 analysis. This table includes the basic kinematic requirements that are enforced on the jet and lepton candidates, and the constraints that are imposed on the global activity in the events (in terms of the hadronic activity $H_T$, the missing transverse momentum $p_T^{\rm miss}$ and various vetoes). We next collect in the left part of Table~\ref{tab:sr} the definition of the most relevant of the 14 signal regions (SRs) of the analysis, as function of the requirements on the number of leptons ($N_\ell$), $b$-tagged jets ($N_b$) and jets ($N_j$). In the right panel of the table, we provide the corresponding expected number of background and signal events as reported by the CMS collaboration.

In the above procedure, a first important point is that signal object candidates are required to satisfy strong isolation criteria. Therefore, the selection ensures the presence of a large number of isolated jets and leptons. Furthermore, given the large number of required $b$-tagged jets, a precise description of the CMS $b$-tagging performance is crucial to obtain an accurate modelling of the search. This represents the major change implemented in the \del\ card dedicated to this search (that is now shipped with the Public Analysis Database of \ma), relative to the default CMS parametrisation in \del. This tuned card allows one to reproduce the efficiencies of the deep neural network $b$-tagging algorithm of CMS (DeepCSV)~\cite{Sirunyan:2017ezt}, whose medium working point configuration is used in the search under consideration.

For the validation of our implementation, we have mimicked the simulation setup used in the CMS-TOP-18-003 four-top analysis, and hence generated SM four-top events at NLO in QCD when all EW contributions are neglected. We have convoluted NLO matrix elements with the NLO set of NNPDF~3.0 parton densities~\cite{Ball:2014uwa}, and the full simulation procedure described above has been employed. As detailed extensively in refs.~\cite{Darme:2020hxc,Fuks:2021wpe}, we have found a good agreement, at least within the theoretical errors on our predictions and the systematical and statistical experimental errors as reported by the CMS collaboration.

\subsection{Alternative recasting: a large $H_T$ analysis}
\label{sec:highHT}

A quite general feature of models with heavy new top-philic particles consists of their collider signature involving top quarks characterised by a significant transverse momentum. In the context of the CMS-TOP-18-003 analysis, it is therefore interesting to consider an extra signal region dedicated to events passing all selection cuts of Table~\ref{tab:preselection}, together with the additional  requirement
\begin{align}
\label{eq:HTcut}
    H_T > 1200 \textrm{ GeV} \ ,
\end{align}
where $H_T$ is the sum of the transverse momenta of all jets.
As we do not have access to any correlation information between the different experimental bins of the $H_T$ spectrum above this threshold, we adopt a conservative approach and combine linearly the associated errors, which corresponds to the worst scenario in terms of correlations. Using data from ref.~\cite{Sirunyan:2019wxt} we therefore extract the following information on the SM expectation and observation,
\begin{align}
    N_{\rm{bkd} + \rm{SM}}  = 6.26 \pm 1.3 \ , \qquad  N_{\rm obs}  = 9\ ,
\end{align}
which exhibits a (very) mild tension.

\begin{figure}[t]
	\centering
	\includegraphics[width=0.65\textwidth]{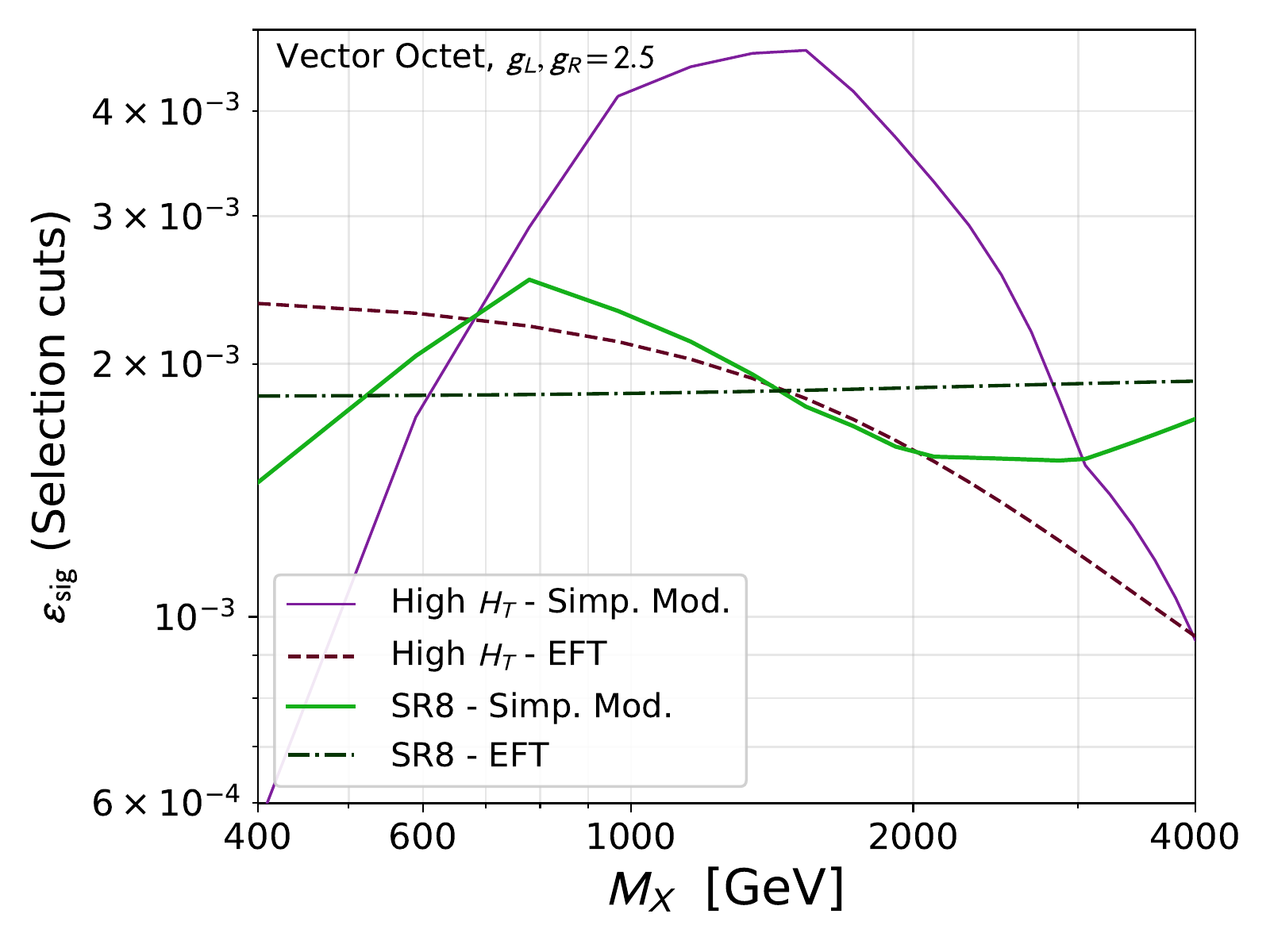}
  \caption{Selection efficiencies $\varepsilon_{\rm sig}$ for the high-$H_T$ signal region (purple lines) and the initial CMS SR8 signal region (green line), as function of the vector-octet mass $M_X=M_{V_8}$ for $\goL = \goR = 2.5$. Results are shown both for the simplified model approach (solid lines) and the matched EFT description (dashed lines).}
  \label{fig:HTEfficienceies}
\end{figure}

We have implemented this new signal region in \ma, and we show in Fig.~\ref{fig:HTEfficienceies} the selection efficiencies  $\varepsilon_{\rm sig}$ obtained in the illustrative example of the colour-octet vector model. Efficiencies are computed both for the new high-$H_T$ signal region, and the SR8 region of the CMS study. This demonstrates the key role potentially played by the high-$H_T$ approach when the top-philic particle lies in the $1-2$~TeV mass window, {\it i.e.}\ when it can be produced on-shell and then decay in a pair of high-$p_T$ top quarks. For increasing and  higher mass values, the off-shell production mode contributes more and more and finally dominates, so that the EFT description (represented by dashed lines in the figure) is more accurate. This can be seen from the convergence of the efficiency curves for masses of $3-4$~TeV. For such large masses, the high-$H_T$ efficiencies are smaller than those associated with the initial region of the CMS analysis, as the signal becomes  more evenly spread across the various $H_T$ bins. When dealing with the original CMS analysis, the EFT efficiencies are relatively close to those computed in the simplified model approach. This indicates that a naive recasting of an EFT interpretation of the CMS results could be performed through a simple correction of the overall signal cross section. Alternatively, this reflects that the current analysis do not leverage the possible impact of on-shell top-philic particles in four-top final states.

\subsection{Numerical procedure}
\label{sec:fullnumerics}

In this subsection, we detail the numerical procedure that we follow to obtain limits on the simplified models presented in Sec.~\ref{sec:model}. We use the chain of high-energy physics tools introduced at the beginning of Sec.~\ref{sec:recast} to simulate 200,000 events for a variety of configurations in each model. We consider scenarios with a coupling value set to either 0.1, 0.3, 1, 3 or 5, and a selection of new physics masses $m_X$ ranging from $100$ GeV to $4$ TeV. The efficiencies for each signal region (including the high-$H_T$ additional SR of Sec.~\ref{sec:highHT}) found by \ma\ are translated into a limit on the maximum signal cross section allowed at 95\% C.L.~for each mass. We interpolate the results to obtain a bound $\sigma^{\rm lim} (\mxt)$ on the signal cross section for any new physics mass $\mxt$, such a bound being then converted into a limit on the new physics couplings as a function of the mass $\mxt$. In practice, this is achieved by solving the implicit equation
\begin{align}
    \sigma(y^{\rm lim}, \mxt) = \sigma^{\rm lim} (\mxt) \ ,
\end{align}
where the cross section value $\sigma(y^{\rm lim}, \mxt)$ is estimated by using a much denser grid in mass and coupling. Such a cross section grid can easily be computed, as it does not require to generate any event or simulate the detector response. The accuracy of this procedure relies on the fact that the selection efficiency varies much more slowly than the cross section as a function of $\mxt$. We derive the error on our estimates from the theoretical uncertainties on the cross section as obtained by varying the higher-order $K$-factor between $1$ and $2$ (see Sec.~\ref{sec:cs}).

\begin{table}
	\begin{center}
		\begin{tabular}{c|cccccccc}
			
			\rule{0pt}{1.25em}
		&SR5 & SR6 & SR7 & SR8 &SR12 & SR13 &SR14 & High $H_T$	\\
	\hline 	\rule{0pt}{1.25em}  $N_{\rm SR}^{\textrm{lim}}$ (LHC) 	& $5$& $11$ & $2.9$ & $7$ &$5$ & $5$ &$4$ & $10$\\[0.2em]
		$N_{\rm SR}^{\textrm{lim}}$ (HL-LHC) &	$58$& $38$ & $25$ & $71$ &$31$ & $26$ &$26$ & $60$ 
		\end{tabular}
		\caption{Maximum number of events populating the most relevant signal regions of the CMS-TOP-18-003 analysis and the high-$H_T$ additional region proposed in this work. We show the observed limits from LHC run~2 data at 95\% C.L., together with the expected projections at the HL-LHC assuming a linear scaling of the number of events and their errors with the luminosity.}
		\label{tab:Nlim}
	\end{center}
\end{table}%

For the evaluation of the EFT limits, we adopt a slightly different approach.  We have simulated 200,000 events for each $P$-conserving operator of the basis \eqref{eq:fullbasis}, once without and once with the interference with the SM contributions. We have also considered the joint impact of any pair of two operators. From those results, we extract, for each SR, two sets of efficiencies $\eNPij{ij}$ and $\eInti{i}$. The former is associated with the $1/\Lambda^4$ contributions of the interference between two (possibly equal) EFT operators ${\cal O}_i$ and ${\cal O}_j$, while the latter is associated with the $1/\Lambda^2$ contributions of the interference between the EFT operator ${\cal O}_i$ and the SM diagrams. The number of signal events $N_{\rm SR}$ populating a specific signal region is then estimated as
\begin{align}
\label{eq:nevent}
    N_{\rm SR} = \mathcal{L}_{\rm CMS} \times  \left( \sum_{ij} \eNPij{ij} \ \sigma^{\nptwo}_{ij} \ \frac{c_i(\tphi) c_j(\tphi)}{\Lambda^4} +  \sum_{i} \eInti{i} \ \sigma^{\Int}_{i} \ \frac{c_i(\tphi)}{\Lambda^2} \right) \ ,
\end{align}
where the sum runs over all EFT operators taken with the corresponding Wilson coefficients $c_i(\tphi)$, and where $\mathcal{L}_{\rm CMS} \equiv 137 \, \textrm{fb}^{-1}$ stands for the LHC run~2 luminosity analysed by the CMS collaboration. The four-top cross sections $\sigma^{\nptwo}_{ij}$ (related to the interference between diagrams involving an operator ${\cal O}_i$ and those involving an operator ${\cal O}_j$) and $\sigma^{\Int}_{i}$ (related to the interference between the SM diagrams and those involving an operator ${\cal O}_i$) are all estimated with {\sc MadGraph5\_aMC@NLO}, although the `diagonal' $\sigma^{\nptwo}_{ii}$ contributions typically dominates over the off-diagonal ones. For a given choice of operators and Wilson coefficients, we can straightforwardly extract a bound on the EFT scale $\Lambda$ by solving the quadratic equation in $\Lambda^2$, \begin{equation}
  N_{\rm SR} =  N_{\rm SR}^{\rm{lim}}\ .
\end{equation}
In this equation, $N_{\rm SR}^{\rm{lim}}$ is the maximum number of signal events, as obtained from \ma, that can populate a given signal region and that is allowed by data at 95\% C.L. The obtained numerical values are tabulated in Table~\ref{tab:Nlim}, both for the observations after the LHC run~2, and for 3~ab$^{-1}$ of proton-proton collisions at the HL-LHC when we assume a linear scaling of the number of events and their errors with the luminosity~\cite{Araz:2019otb}. 

\section{Results}
\label{sec:results}

\begin{figure}[t]
\centering
\subfloat[]{%
\includegraphics[width=0.47\textwidth]{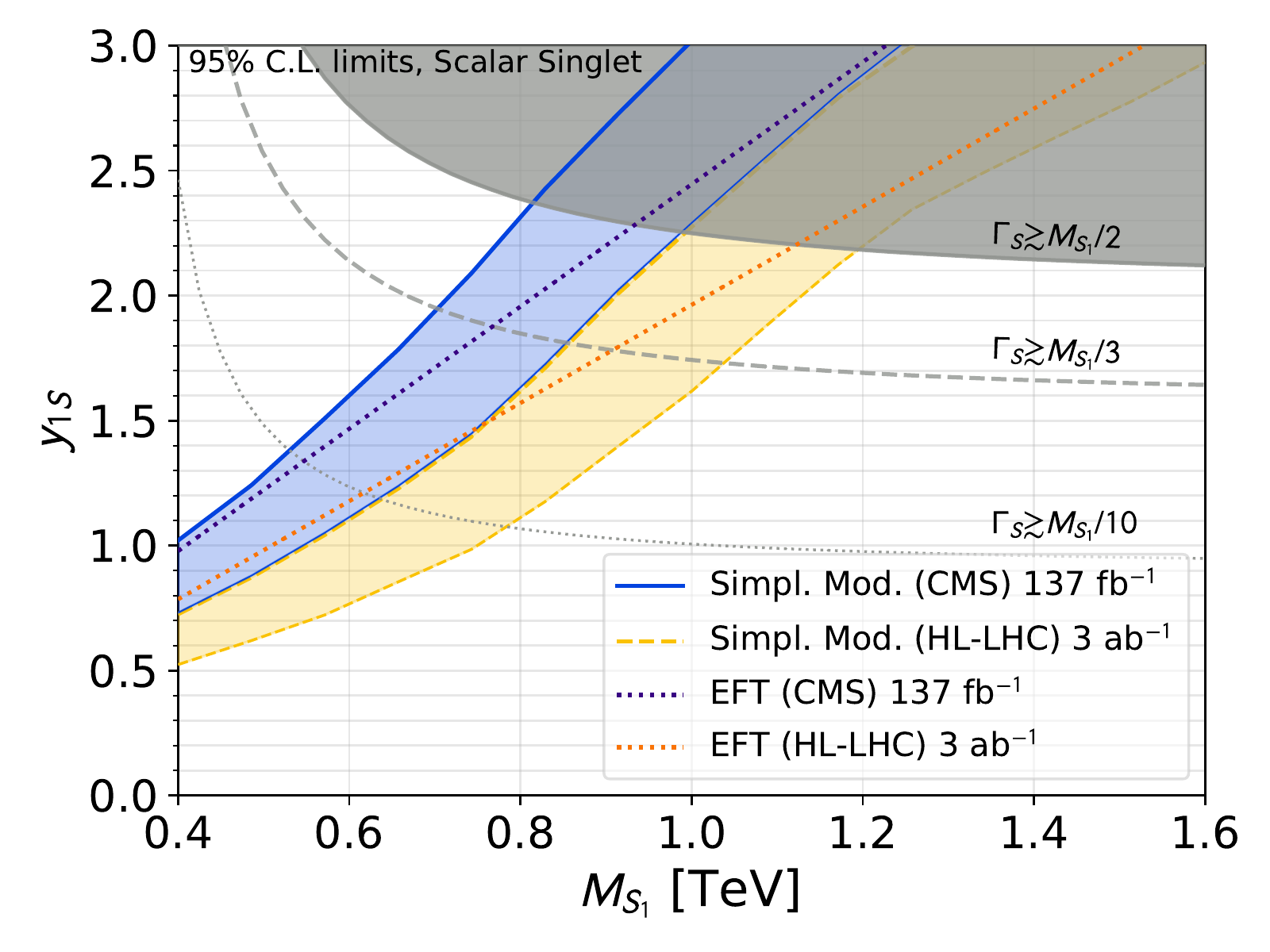}
}%
\hspace{0.02\textwidth}
\subfloat[]{%
\includegraphics[width=0.47\textwidth]{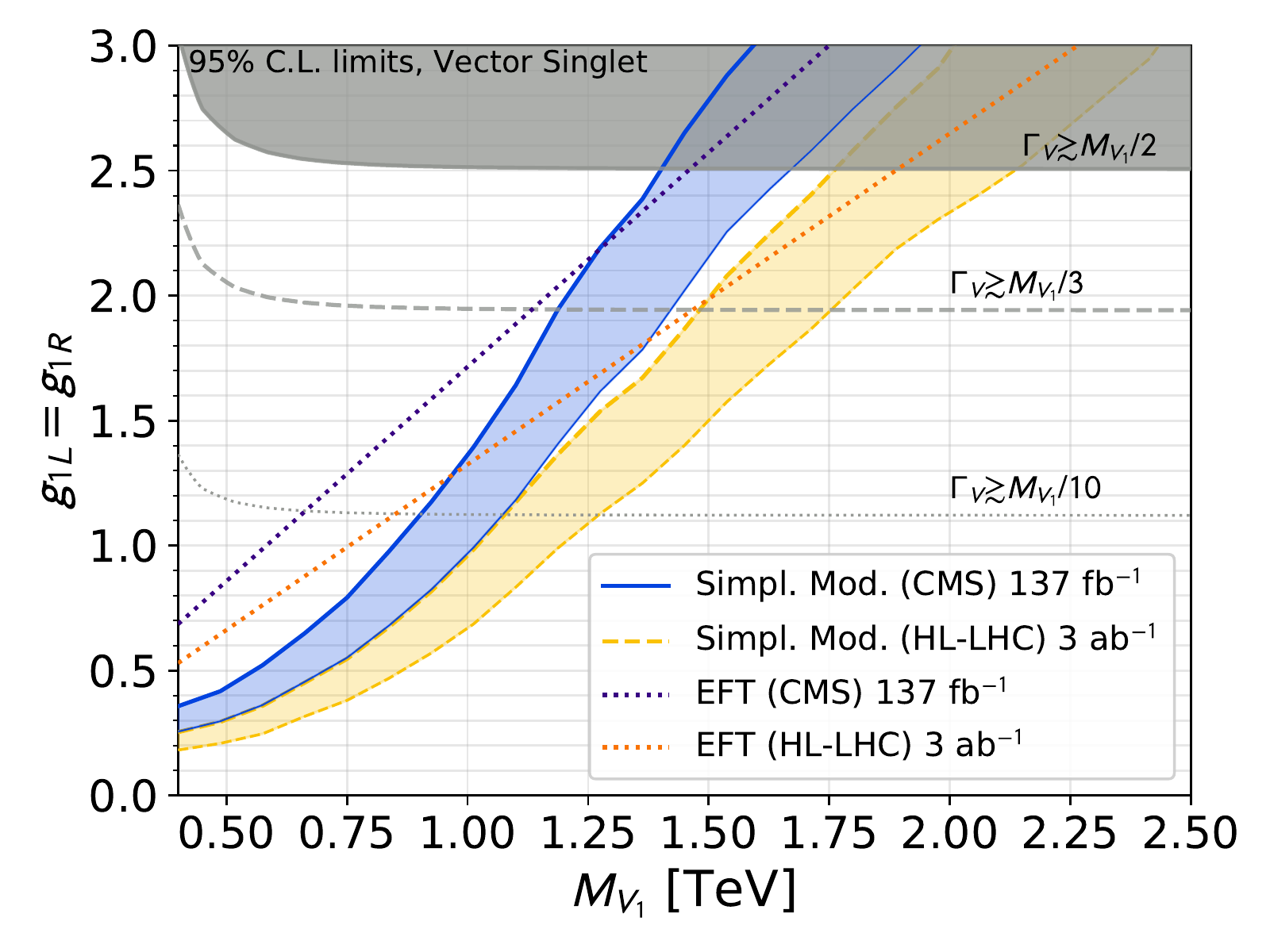}
}%
\caption{95\% C.L.~exclusions for two top-philic  simplified models, presented in mass {\it versus} coupling planes and derived from the CMS-TOP-18-003 analysis (including the additional signal region of Sec.~\ref{sec:highHT}). We consider (a) $\ysS$ as function of $M_{\sing}$ for a top-philic scalar singlet simplified model, and (b) $\gsL = \gsR$ as function of $M_{\vs}$ for a top-philic vector singlet simplified model. The blue lines correspond to the current CMS limit and the golden lines represent our projection for the HL-LHC with an integrated luminosity of $3 \, \textrm{ab}^{-1}$, the regions above the curves being excluded in both cases. The dotted lines are the EFT-derived limits, while the solid and dashed lines represent the simplified model results. The EFT limits include a $K$-factor of $1$, while the simplified model results are shown as bands obtained by varying the theoretical $K$-factor between $1$ (top line) and $2$ (bottom line). The grey regions in the top-right corners represent the parameter space region in which the width of the top-philic particle is too large, making our approach unreliable.}
\label{fig:ylimsinglet}
\end{figure}

Using the procedure detailed in Sec.~\ref{sec:recast}, we have compared the limits obtained within the EFT formalism with those arising from the simplified model approach. In the process, we have obtained the most up-to-date limits from four-top probes on scalar and vector octet models as described in Sec.~\ref{sec:model}. 

In Fig.~\ref{fig:ylimsinglet}a, we show the limits obtained in the case of a top-philic scalar singlet model when the results of the CMS-TOP-18-003 analysis are reinterpreted, after including the additional signal region of Sec.~\ref{sec:highHT}. We present the bounds as 95\% C.L.~exclusion contours in a mass {\it versus} coupling plane $(M_{\sing}, \ysS)$, including the uncertainty on NLO corrections as discussed in Sec.~\ref{sec:cs} ({\it i.e.}~by varying the $K$-factor between $1$ and $2$). The considered CMS search can probe this scenario for new physics masses ranging up to about $0.8$ TeV for a coupling $\ysS = 2.25$. We expect a small improvement at the HL-LHC, which pushes the limit to roughly $1.1$ TeV for the same value of the coupling. We have overlaid in the figures the corresponding EFT limits, that we extract via the matching conditions of Table~1. These limits therefore correspond to those that one would obtain by translating an EFT limit into a constraint on the couplings appearing in vector and scalar singlet models, without accounting for the presence of any new on-shell resonance. Fortuitously, the ``naive'' EFT approach reproduces relatively well the simplified model limits. As can be seen in Fig.~\ref{fig:CSscalarInt}, this  is due to the fact that both the simplified model and the EFT approach lead to the same cross section in the coupling and mass range to which the LHC is sensitive. This however happens only in a regime where an EFT description of the complete simplified model is beyond applicability, as pointed out in Sec.~\ref{sec:cs}.
In Fig.~\ref{fig:ylimsinglet}b, we show the same bounds, but for the case of a top-philic vector singlet model. Although the range of coupling values accessible at the LHC is roughly similar to the scalar case, the agreement between the EFT and the simplified model is worse at low masses. This can be traced back to the fact that EFT predictions significantly underestimate the cross section for the relevant coupling range (see Fig.~\ref{fig:CSvectorInt}).

\begin{figure}[t]
\centering
\subfloat[]{%
\includegraphics[width=0.47\textwidth]{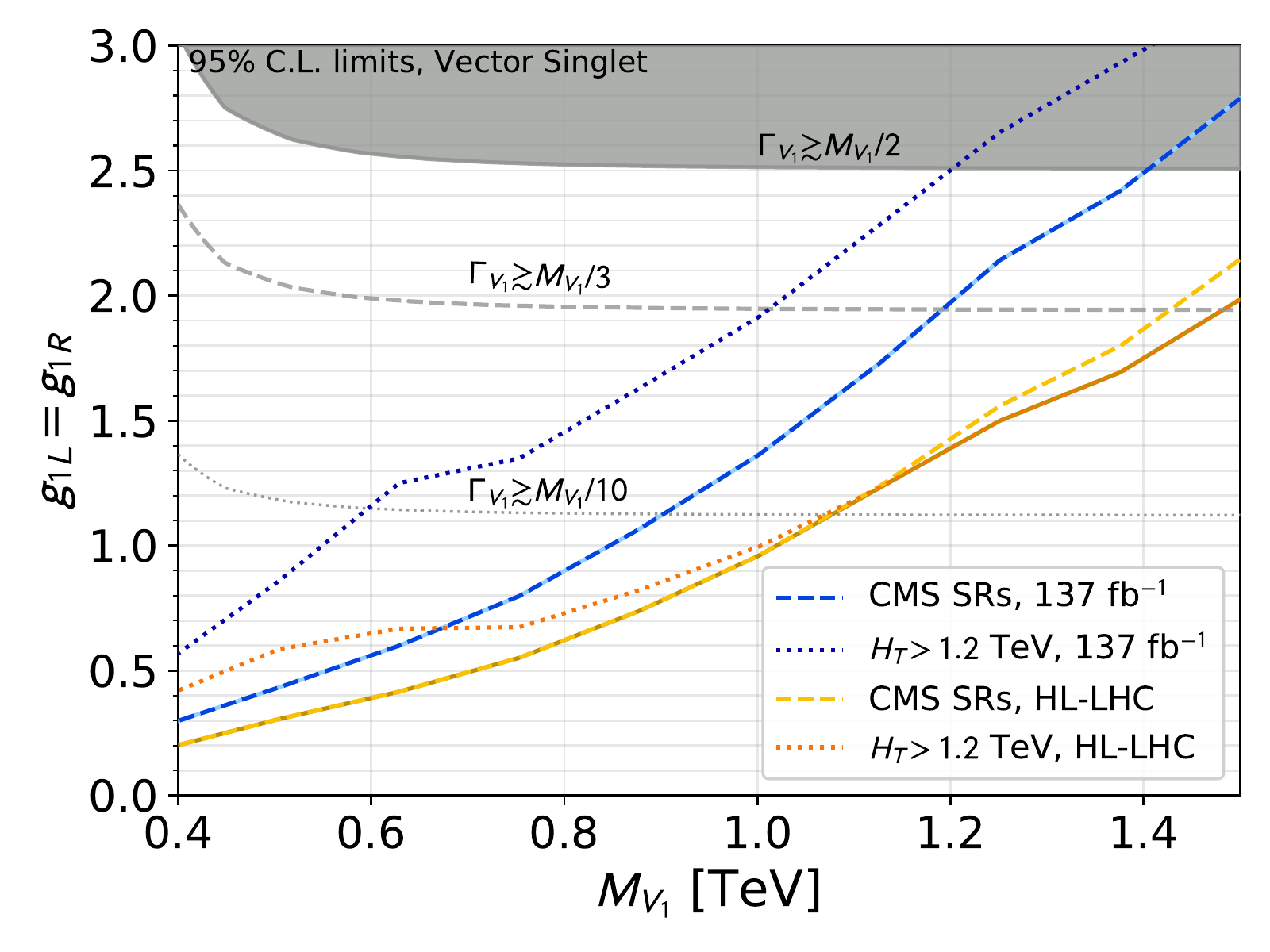}
}%
\subfloat[]{%
\includegraphics[width=0.47\textwidth]{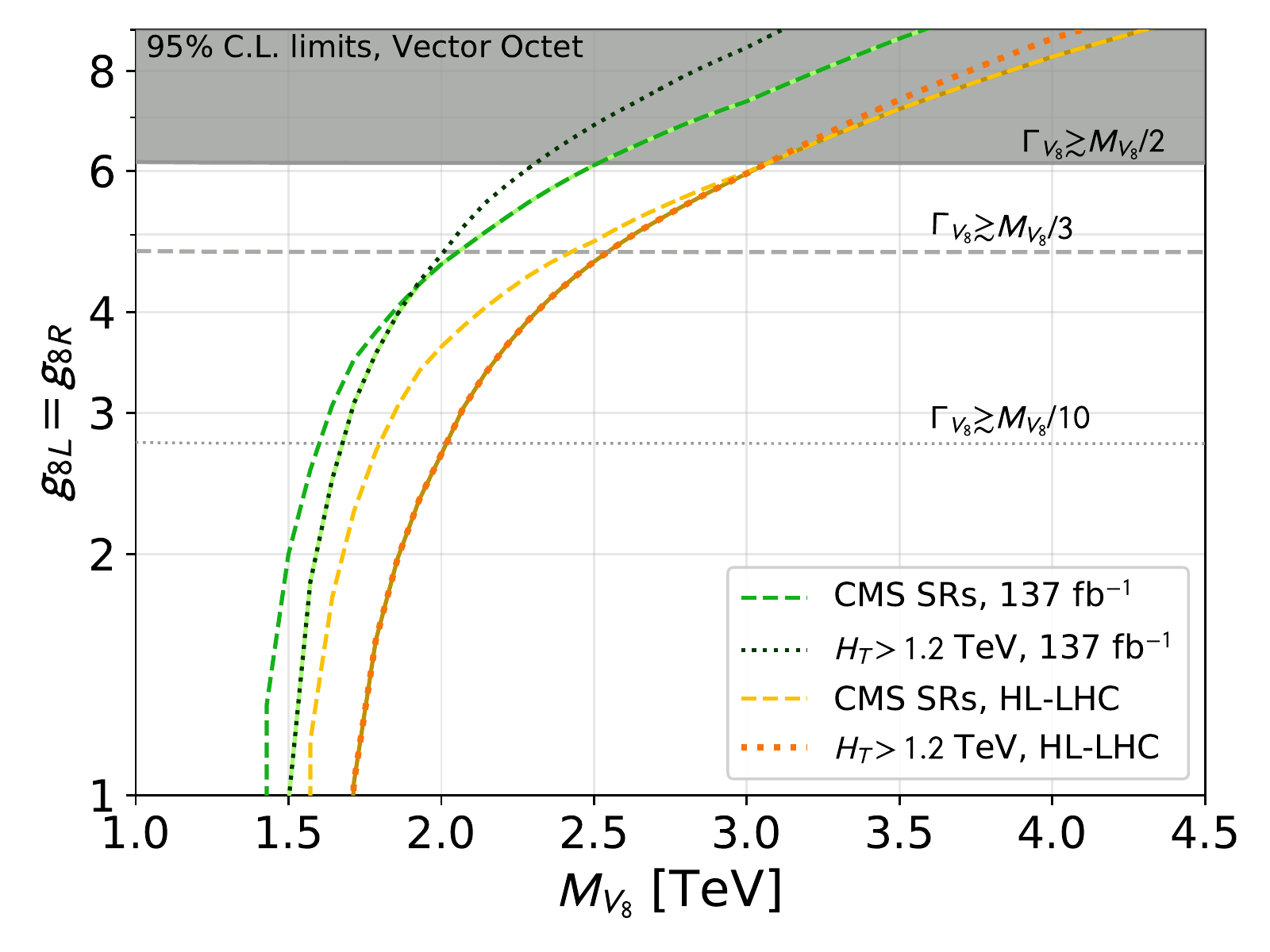}
}
\caption{95\% C.L.~exclusions for two top-philic simplified models, presented in mass {\it versus} coupling planes and derived from the CMS-TOP-18-003 analysis (dashed lines) and when relying on the additional signal region of Sec.~\ref{sec:highHT} (dotted lines). We consider (a) $\gsL = \gsR$ as a function of $M_{\vs}$ for a vector singlet simplified model, and (b) $\goL = \goR$ as a function of $M_{\vo}$ for a vector octet model. We include results extracted from LHC run~2 data and our projections for the HL-LHC, the regions above the curves being excluded in both cases. The $K$-factor is fixed at 1. The grey exclusions represent the parameter space regions in which the width of the top-philic particle is too large, making our approach unreliable.}
\label{fig:comparison}
\end{figure}

In Fig.~\ref{fig:comparison}, we focus on the case of the colour-singlet (a) and colour-octet (b) vector simplified models, the results  distinguishing this time the constraints arising from the original CMS-TOP-18-003 search (dashed lines) and those originating from the high-$H_T$ approach presented in Sec.~\ref{sec:highHT} (dotted lines). We find that the high-$H_T$ approach dominates the limit  between $1$ and $3$ TeV. This can be traced back to the predominance of on-shell production in this mass range, which in turn ensures that the four final-state top quarks accumulate a large amount of transverse momentum. All previous limits that are based on the reinterpretation of the Standard Model analysis are thus likely to underestimate the LHC sensitivity in this mass range. Furthermore, we observe that recasting the standard CMS search is sufficient for the largest and smallest masses examined in our analysis. In these cases, on-shell production is either sub-dominant (high masses) or the top-philic particle is simply not massive enough to yield high-$p_T$ objects. Altogether, we find that an ``on-shell'' analysis such as the high-$H_T$ approach presented above is particularly well-suited to seek most new top-philic resonances at the HL-LHC, and is almost always the best approach when studying octet states. From now on, we only present the dominant limits.

\begin{figure}[t]
\centering
\subfloat[]{%
\includegraphics[width=0.47\textwidth]{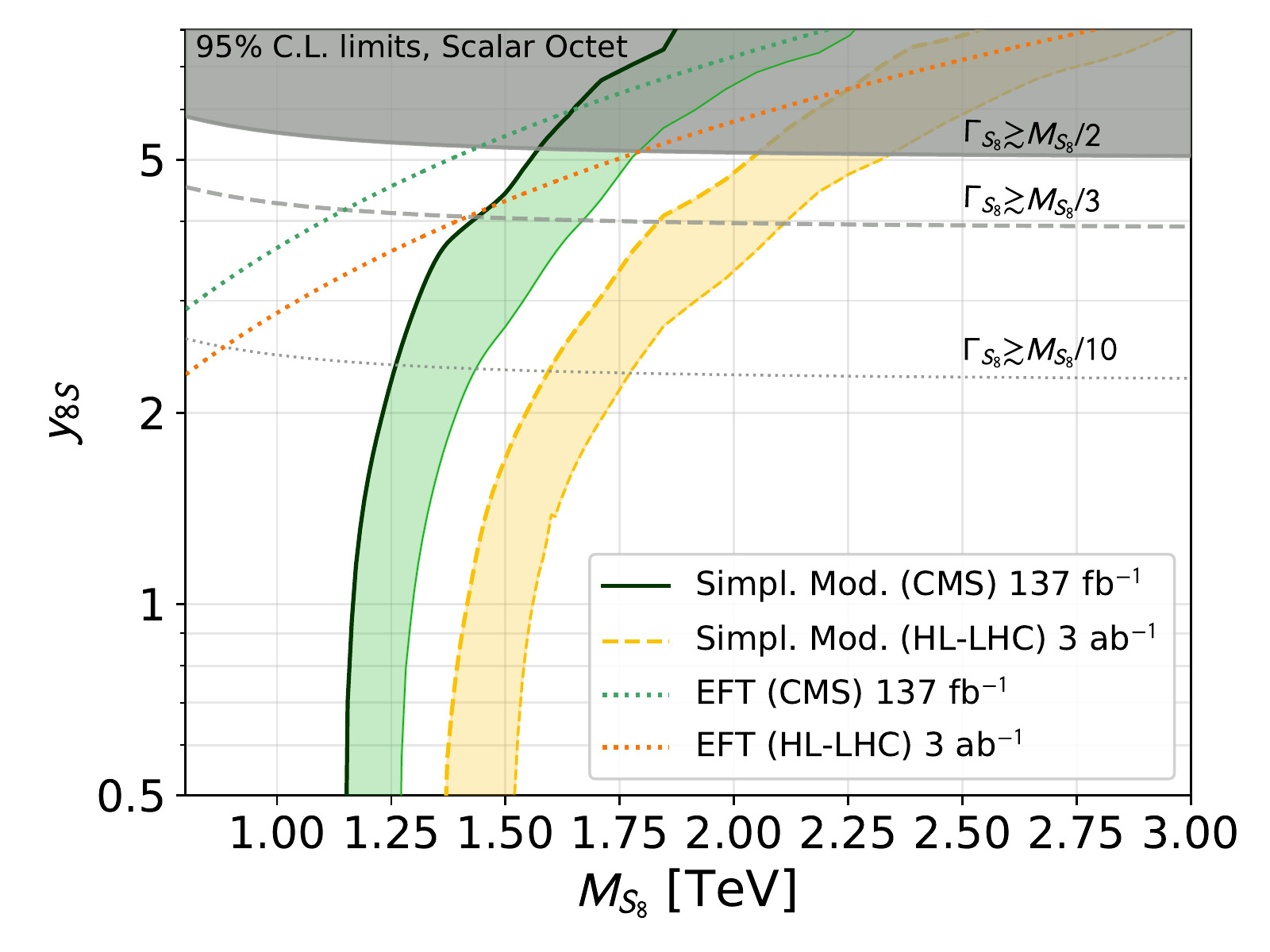}
}%
\hspace{0.02\textwidth}
\subfloat[]{%
\includegraphics[width=0.47\textwidth]{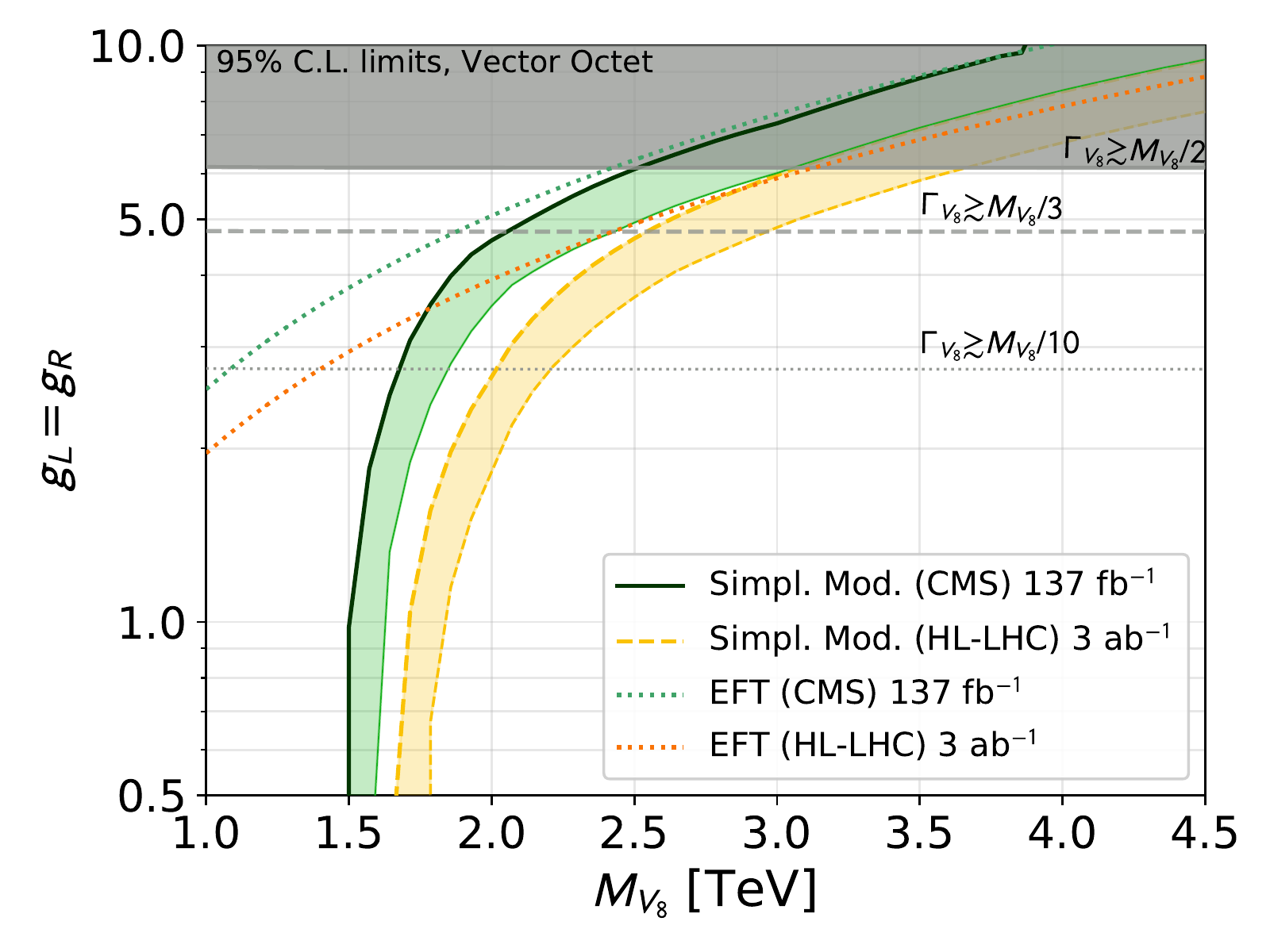}
}%
\caption{95\% C.L. exclusions for two top-philic simplified models, presented in mass {\it versus} coupling planes and derived from the CMS-TOP-18-003 analysis (including the additional high-$H_T$ signal region of Sec.~\ref{sec:highHT}). We consider (a) $\yoS$ as function of $M_{\oct}$ for a top-philic scalar octet simplified model and (b) $\goL = \goR$ as function of $M_{\vo}$ for a top-philic vector octet simplified model. Our results include the reinterpretation of the current CMS results (green lines) and their extrapolations at the HL-LHC (golden lines). The dotted lines represent the EFT-derived limits, while the solid and dashed lines represent the simplified model results. For the EFT limits, we use a $K$-factor of $1$, while the simplified model ones are shown as bands obtained by varying the theoretical $K$-factor between $1$ (top line) and $2$ (bottom line). The grey regions in the top-right corner represent the parameter space regions in which the width of the top-philic particle is too large, making our approach unreliable.}
\label{fig:ylimoctet}
\end{figure}

In Fig.~\ref{fig:ylimoctet}a, we focus on the case of the scalar octet. As expected, the simplified model can be robustly excluded for top-philic state masses lying below about $1.25$ TeV, even for very small couplings. This is due to the QCD-induced pair-production mechanism relevant for octet states that leads to a very large cross section for masses smaller than about 1~TeV. Our results improve those derived in the earlier analysis of ref.~\cite{Darme:2018dvz}, which was based on the reinterpretation of the results of the LHC for a smaller dataset, including only a luminosity of $35.9$ fb$^{-1}$. Moreover, in this particular pair-production-driven regime, one can confidently consider that the aggressive limit obtained when using a $K$-factor of $K=2$ is valid, as explained and computed in ref.~\cite{Darme:2018dvz}. We finally investigate the case of a vector octet top-philic particle in Fig.~\ref{fig:ylimoctet}(b). Here, the lower bound for the exclusion is pushed to about $1.6$ TeV. At larger coupling values, the limit extends up to around $3$ TeV, even before leading to an unacceptably large width for the top-philic resonance. We find that the EFT approach is particularly suitable for such a heavy-state regime, leading to an excellent agreement for top-philic masses above $2.25$ TeV. Importantly, the limits that we obtain, for both the scalar and the vector octet cases, rely on the high-$H_T$ analysis for all masses below 3~TeV. 

\begin{figure}[t]
\centering
\subfloat[]{%
\includegraphics[width=0.47\textwidth]{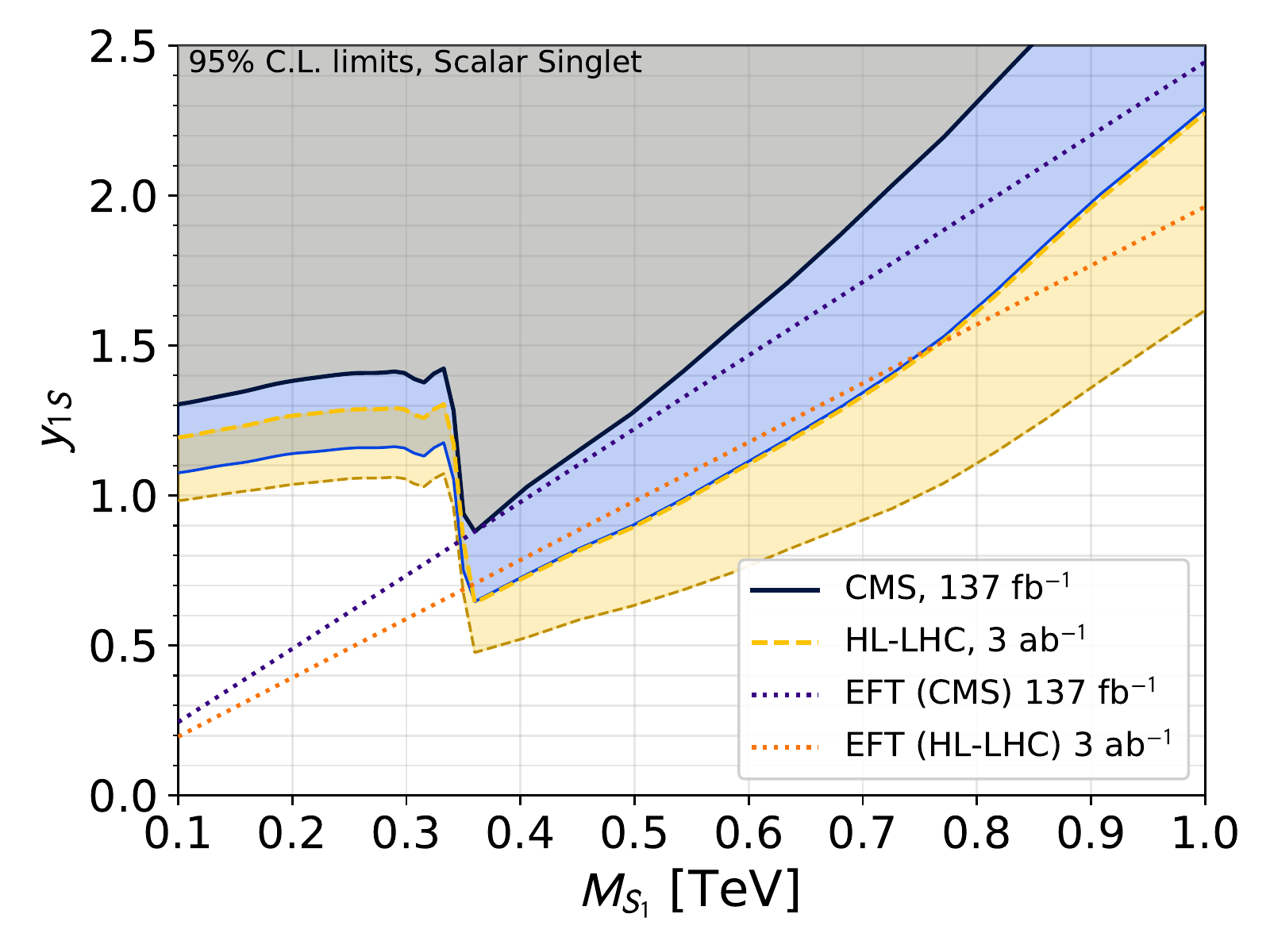}
}%
\hspace{0.02\textwidth}
\subfloat[]{%
\includegraphics[width=0.47\textwidth]{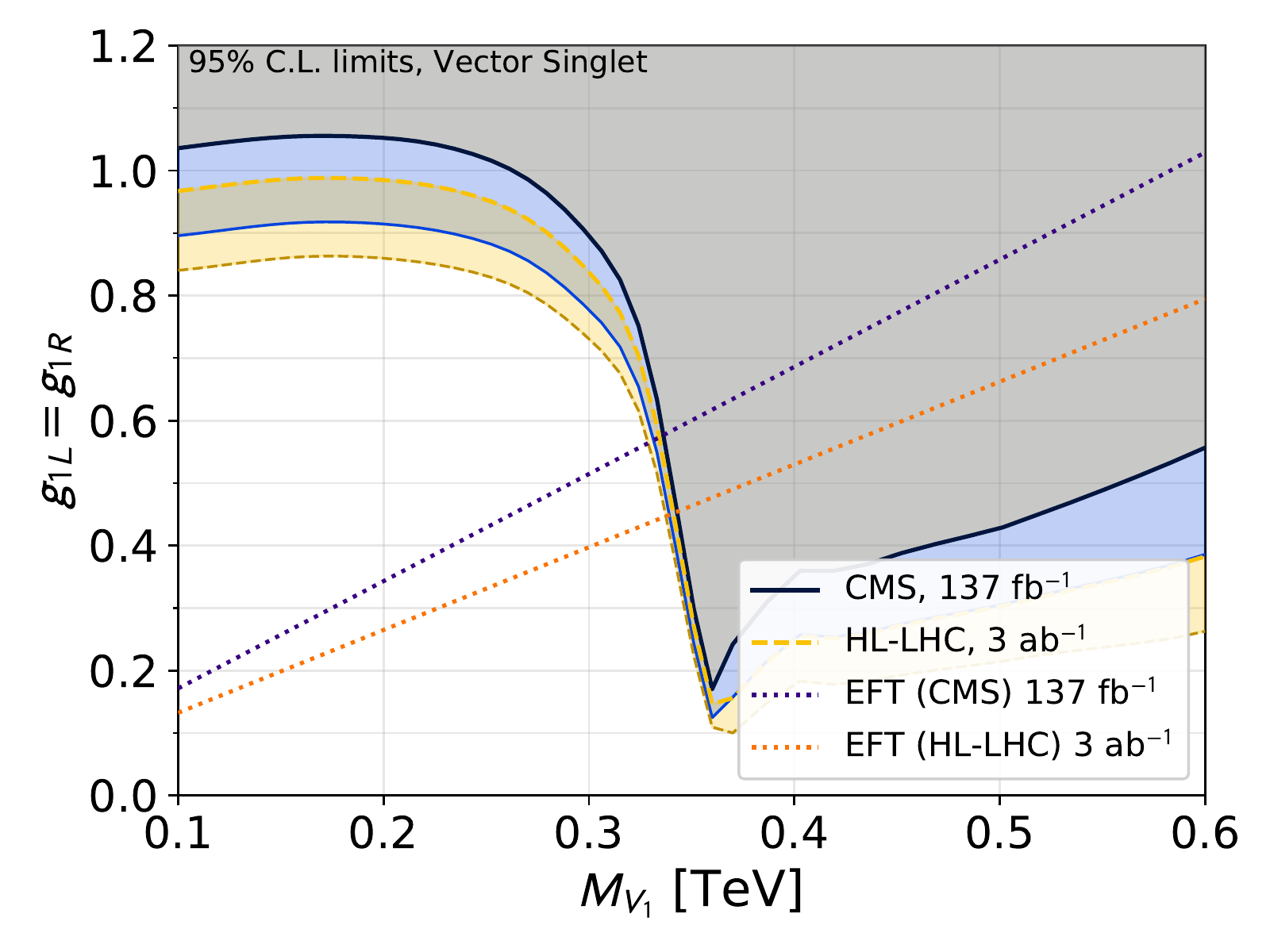}
}%
\caption{Same as Fig.~\ref{fig:ylimsinglet}, but for a zoom on the $[100, 1000]$ GeV mass range. The EFT limits are shown with thin dashed lines to emphasise that the EFT approach is not grounded in this parameter space region.
}
\label{fig:ylimlowmass}
\end{figure}

We close this section by commenting further on the phenomenology associated with the low mass region. When the resonance is below the di-top threshold, four-top production from a pair of off-shell top-philic states dominates. However, the scaling of the top-philic state propagator is completely different from the one assumed when computing the EFT limit. Instead of the $1/M^2$ behaviour that would be expected for a large top-philic state mass $M$, it indeed exhibits a $1/s$ behaviour, with $s$ being typically around the invariant mass of the di-top system. We show in Fig.~\ref{fig:ylimlowmass} the limits that we obtain by reinterpreting the CMS-TOP-18-003 results when the top-philic state mass varies in the $100$ to $1000$ GeV mass range, both for the scalar singlet and vector singlet simplified models. As expected from the above argument, the naive EFT approach overestimates the limits in this range. This is emphasised by the results shown in Sec.~\ref{sec:4tCS}, from which we can observe that the signal efficiency would be drastically reduced after accounting for the fact that the EFT is used in a regime where it should not be used (see Fig.~\ref{fig:EffEFTScale}). Additionally, it is clear that the topology of the final state resembles closely the one expected from SM Higgs-induced four-top production, as verified directly by the CMS collaboration~\cite{Sirunyan:2019wxt}. In this regime, we therefore consider the boosted decision tree (BDT) analysis of CMS~\cite{Sirunyan:2019wxt} and extract limits directly from the measured signal cross section 
\begin{align}
    \sigma_{4t}^{\rm NP + \rm SM} = 12.6^{+5.8}_{-5.4}~{\rm fb} \ .
\end{align}
In order to derive projections for the HL-LHC, we use the last three BDT signal regions SR15 to SR17, that lead to 
\begin{align}
    N_{\rm{bkd} }  = 8.1 \pm 1.3  \ , \qquad N_{\rm{SM}}  =8.2 \pm 3.1  \ , \qquad  N_{\rm obs}  = 14\ .
\end{align}
Assuming the same background and BDT efficiencies, we extract from the CMS findings an upper limit on the full (SM plus new physics) cross section of $\sigma^{\rm NP + \rm SM}_{95, \rm HL} = 23~{\rm fb} $ at 95\% C.L. As this is compatible with the current theoretical uncertainties on the SM four-top production cross section, we add the latter in quadrature with the projected experimental errors to obtain a global error on the SM background uncertainties. We then obtain a rough limit at 95\% C.L on the new physics contributions to the four-top cross section of $\sigma^{\rm NP}_{95\%, \textrm{HL}} \sim 8~{\rm fb}$ assuming that the observed fluctuation at the end of run~2 will be reproduced at HL-LHC, or of about 10~fb otherwise. The corresponding moderate improvements of the limits at HL-LHC corresponding to the former case are shown in Fig.~\ref{fig:ylimlowmass}(a), and then reflect the important role of the theoretical and systematic errors in the projections that consist of their limiting factor.

The same simplified model including a scalar singlet was also considered directly in refs.~\cite{Sirunyan:2019wxt,Alvarez:2016nrz}. Our limits are comparable with those derived in those works. We additionally consider, in Fig.~\ref{fig:ylimlowmass}(b), the limits that are obtained for a vector singlet simplified model. Compared to refs.~\cite{Alvarez:2016nrz,Sirunyan:2019wxt,Alvarez:2020ffi}, we use a `vectorial' interaction of the top-philic state with the top quark, so that the cross section predictions do not exhibit any chirality-flip enhancement at low mass. While the  vector model with purely chiral interactions and additional flavour-violating interactions of  ref.~\cite{Alvarez:2020ffi} allows for a good fit to ATLAS data,  we have not directly examined this option in our study, any flavour-violating coupling being absent from our beyond the SM parametrisation.

\section{Conclusions and outlook}\label{sec:concl}

In this work, we have considered the status and outlined the prospects for constraining the existence of  top-philic resonances by using four-top probes at the LHC. Motivated by  the growing interest in applying the SMEFT interpretational framework to heavy new physics searches, we have compared the predictions for several simplified models with the  expectations based on the corresponding  EFT descriptions. In order to cover a wider range of simplified model candidates, some of which including $SU(2)_{\rm L}$-symmetry breaking terms, we have parametrised four top-quark observables in terms of operators in the $SU(3)_{\rm C}\times U(1)_{\rm em}$ basis. The matching coefficients between the simplified model descriptions and the EFT basis have been determined analytically and validated numerically. In passing, we have found that the interference of EFT amplitudes with EW-induced SM contributions, which has been so far neglected in the EFT analyses, provides a sizeable component of the total EFT cross section. Due to this, the available NLO in QCD calculations in the EFT cannot be representative of the regimes where the interference with the SM is the dominant contribution. The pattern of the C-NLO corrections is rather intricate already in the SM, with significant cancellations taking place between contributions at different orders. As a result, applying a  multiplicative ``$K$-factor'' similar to the SM one to new physics scenarios cannot be considered reliable in all EFT regimes. We have therefore chosen to keep the central values at LO and associate a large theoretical uncertainty, between a factor of $1$ and $2$, when presenting our limits.

Our findings indicate that at the LHC and in the majority of the cases the EFT approach significantly underestimates both the signal cross section and the final limits on the new top-philic particle couplings. This originates from resonance pair-production contributions that dominate the cross section. The signal efficiencies of both approaches are nonetheless relatively similar in the case of the CMS analysis that we have reinterpreted in this work, which has been optimised to the SM signal. The updated coupling-independent limits arising from QCD-induced pair production of color octets are $M_{\oct} \gtrsim 1.25$ TeV and $M_{\vo} \gtrsim 1.6$ TeV in the scalar and vector cases respectively. One possibility to make the EFT interpretation more accurate would therefore be to invoke specific mechanism to reduce on-shell contributions to four-top production, as for instance when new top-philic particles decay into new states beyond the SM. 

Finally, we have derived our constraints by relying on the combination of two analyses: a pure reinterpretation of the CMS four-top searches based on the existing signal regions, and an additional one defined by recasting their data in the large $H_T$ region. While not completely optimised due to missing information (like \eg\ background correlations), the latter typically dominates the limits for masses between $1$ and $2$ TeV. This can be traced back to large differences in kinematics between the Standard Model background and the new physics signal, particularly regarding the total transverse momentum of the final state. These results emphasise the need to continue dedicated top-philic search strategies by ATLAS and CMS during the LHC run~3, and in the up-coming High-Luminosity phase of the LHC.

\subsubsection*{Acknowledgments}
We are thankful to Eleni Vryonidou and Gauthier Durieux for comments on the manuscript and many enlightening discussions. LD and FM are supported by the INFN “Iniziativa Specifica” Theoretical Astroparticle Physics (TAsP-LNF) and 
QFT@Colliders, respectively. FM acknowledges the support of the F.R.S.-FNRS under the Excellence of Science EOS be.h project n. 30820817.

\newpage

\appendix

\addcontentsline{toc}{section}{Appendices}

\section{Effective Field Theory basis and matching}
\label{sec:matching}

 When dealing with four-top production in the $SU(3)_{\rm C} \times U(1)_{\rm em}$ basis, the standard SMEFT basis is expectedly overcomplete. We review in this appendix the reduction process and our final choice for the EFT basis considered in this work.
 
The first step is the Fierz decomposition of a bispinor $\fa \fab$, where $i$ and $j$ are for the moment generic indices. After including a global minus sign as we deal with anti-commuting fermion, we get
\begin{align}
    \fa \fab = - \frac{1}{4} \left( \fab \fa \mathbb{1} +  \fab  \gamma^5 \fa  \ \gamma^5 + \fab \gamma^\mu \fa \ \gamma_\mu - \fab \gamma^\mu\gamma^5 \fa \ \gamma_\mu \gamma^5  + \frac{1}{2} \fab \sigma^{\mu \nu} \fa \sigma_{\mu \nu} \right) \ .
\end{align}
In this expression, we have used the standard notation $\sigma^{\mu \nu} = \frac{i}{2} (\gamma^\mu \gamma^\nu-\gamma^\nu \gamma^\mu)$. In order to properly handle the associated colour structure, this last equation needs to be combined with $SU(3)$ algebra relations allowing for the permutation of the colour indices,
 \begin{align}
     T^A_{kj} T^A_{il} = \frac{4}{9} \delta_{ij} \delta_{kl} - \frac{1}{3} T^A_{ij} T^A_{kl} \\
      \delta_{kj} \delta_{il} = \frac{1}{3} \delta_{ij} \delta_{kl} +2 T^A_{ij} T^A_{kl} \label{eq:fierzsu3del}\ .
 \end{align}
Combining both relations, it is possible to systematically apply a Fierz transformations to any four top operator from the list below, where we have omitted any Lorentz and colour index for clarity,
\begin{align}
    (\Sop) (\Sop), \ (\Sop) (\Pop) , \ (\Pop) (\Pop) , \ (\Vop) (\Vop) ,\\
    (\Vop) (\Aop), \ (\Aop) (\Aop), \\
    (\Top) (\Top), \ (\Top) (\DTop) \ ,
\end{align}
together with their colour-octet equivalent in which the trivial colour contractions $\delta_{ij}$ are replaced by $T^A_{ij}$ ones involving an $SU(3)$ generator.

From the above procedure, we  reduce the whole set of operators to a basis comprising a restricted set of six operators,
 \begin{align}
\oRR&=  \bar{t}_R \gamma^{\mu} t_R   ~\bar{t}_R \gamma_{\mu} t_R \\
	\oLL&=  \bar{t}_L \gamma^{\mu} t_L   ~\bar{t}_L \gamma_{\mu} t_L \\
	\oLR &=  \bar{t}_L \gamma^{\mu} t_L   ~\bar{t}_R \gamma_{\mu} t_R \\	
 \oTaLR &=  \bar{t}_L T^A \gamma^{\mu} t_L   ~\bar{t}_R T^A \gamma_{\mu} t_R \\
\os&=  \bar{t}  t   ~\bar{t}t \\
\oTas &=  \bar{t} T^A  t   ~\bar{t} T^A t  \ .
\end{align}
To be general, two additional operators could arise,
 \begin{align}
\ops&=  \bar{t}  t   ~\bar{t} (i \gamma^5) t \\
\oTaps &=  \bar{t} T^A  t   ~\bar{t} T^A  (i \gamma^5)  t  \ ,
\end{align}
which, however, do not typically appear in any of the simplified models probed in this work. The coefficients of the decomposition of the other operators onto this basis are given in Table~\ref{tab:Fierzdecomp}. 
 \begin{table}
	\begin{center}
		\begin{tabular}{c|c|c|c|c|c|c|c|c}
			\rule{0pt}{1.25em}
		& $\os$ & $\oTas$ & $\oLL$  &  $\oRR$ &  $\oLR$ & $\oTaLR$  & $\ops$ & $\oTaps$ \\[0.2em]
			\hline 	\rule{0pt}{1.25em}
			$\bar{t} \gamma^5 t \ \bar{t} \gamma^5 t$ & $1$ & / & /  &  / & $2/3$ & $4$ & /  &  / \\[0.2em]
			$ T^A T^A  \ \bar{t} \gamma^5 t  \ \bar{t} \gamma^5 t$ & / & $1$ & /  &  / & $8/9$ & $-2/3$ & /  &  /\\[0.2em]
        	\hline
			\rule{0pt}{1.25em}
        	$ T^A T^A  \ \bar{t}_L \gamma t_L  \  \bar{t}_L \gamma t_L $ & / & / & $1/3$  & / & / & /& /  &  / \\[0.4em]
        	$ T^A T^A  \ \bar{t}_R \gamma  t_R  \  \bar{t}_R \gamma  t_R $ & / & / & / &  $1/3$ & / & / & /  &  /\\[0.4em]
        	  \hline
			\rule{0pt}{1.25em}
		$\bar{t} \sigma t \ \bar{t} \sigma t$ & $-20/3$ & $-16$ & /  &  / & $-28/3$ & $-8$ & /  &  /\\[0.2em]
$ T^A T^A  \ \bar{t} \sigma t \ \bar{t} \sigma t \ $ & $-32/9$ & $-4/3$ & /  &  / & $-16/9$ & $-20/3$ & /  &  /\\[0.2em]
		$\bar{t}  \sigma (i \gamma^5) t \ \bar{t} \sigma t$ & / & / /  &  / & / & / &/ & $-20/3$ &  $-16$\\[0.2em]
$ T^A T^A  \ \bar{t} \sigma  (i \gamma^5) t \ \bar{t} \sigma t \ $ & / & /& /  &  / & / & / & $-32/9$  &  $-4/3$\\[0.2em]

		\end{tabular}
		\caption{Decomposition of the most general over-complete set of dimension-six four-top EFT operators into a basis comprising the operators~\eqref{eq:OeffEW} and~\eqref{eq:OeffEWbreaking}.} 
		\label{tab:Fierzdecomp}
	\end{center}
\end{table}%

In order to compare the above basis with the one that could be expected from the SMEFT in the Warsaw basis~\cite{Grzadkowski:2010es}, it is useful to recall the Fierz rule that is associated with the $SU(2)$ algebra (with $i,j,k,l$ indices being $SU(2)$ fundamental indices and $\tau^I$ denoting the Pauli matrices),
\begin{align}
    \tau^I_{ij} \tau^{I}_{kl} =  2 \delta_{il} \delta_{jk} -  \delta_{ij} \delta_{kl} \ .
\end{align}
Specifying to third generation quarks only, the left-handed ``triplet'' and ``singlet'' operators in the Warsaw basis read
\begin{align}
\mathcal{Q}_{qq}^{(3)} &= \bar{q}_L \tau^I \gamma^{\mu} q_L   ~\bar{q}_L \tau^I \gamma_{\mu} q_L \\
\mathcal{Q}_{qq}^{(1)} &= \bar{q}_L \gamma^{\mu} q_L   ~\bar{q}_L \gamma_{\mu} q_L \ 
\end{align}
and can be decomposed into
\begin{align}
    \mathcal{Q}_{qq}^{(3)} &= \oLL + \bar{b}_L \gamma^{\mu} b_L   ~\bar{b}_L \gamma_{\mu} b_L + 4 \bar{b}_L \gamma^{\mu} t_L   ~\bar{t}_L \gamma_{\mu} b_L - 2 \bar{b}_L \gamma^{\mu} b_L   ~\bar{t}_L \gamma_{\mu} t_L \nonumber \\
    \mathcal{Q}_{qq}^{(1)} &= \oLL + \bar{b}_L \gamma^{\mu} b_L   ~\bar{b}_L \gamma_{\mu} b_L + 2 \bar{b}_L \gamma^{\mu} b_L   ~\bar{t}_L \gamma_{\mu} t_L \ .
\end{align}
Similar relations can be derived in the case of colour-octet operators, as the decomposition above only focuses on the $SU(2)$ structure. All those operators are thus clearly redundant with $\oLL$ as regards to the pure four-top phenomenology (with similar trivial relations for the right-handed operators). As an additional cross-check, one can also verify that the SMEFT relations used in ref.~\cite{Grzadkowski:2010es} such as
\begin{align}
   \bar{q}_L T^A \tau^I \gamma^{\mu} q_L   ~\bar{q}_L T^A \tau_I \gamma^{\mu} q_L  = \frac{1}{4}  \mathcal{Q}_{qq}^{(3)} + \frac{1}{12}\mathcal{Q}_{qq}^{(1)}  \ ,
\end{align}
lead to the same pure four-top result as found in Table~\ref{tab:Fierzdecomp}\footnote{This relation can also be obtained directly by using the well-known Fierz relation (see  \eg\ ref.~\cite{Grzadkowski:2010es}) $\bar{t}_{L,i} \gamma^{\mu} t_{L,j}   ~\bar{t}_{L,k} \gamma_{\mu} t_{L,l}  = \bar{t}_{L,i} \gamma^{\mu} t_{L,k}   ~\bar{t}_{L,j} \gamma_{\mu} t_{L,l} $, where $i,j,k,l$ keep tracks of the colour indices, along with the $SU(3)$ relation~\eqref{eq:fierzsu3del} to re-arrange these indices canonically.},
\begin{align}
\label{eq:OctetlefttoSinglet}
\oTaLL = \frac{1}{3} \oLL \ .
\end{align}

In the framework defined by the LHC Top Working Group~\cite{AguilarSaavedra:2018nen} which includes also the interactions with other quarks, the operators with left-handed tops are defined as\footnote{We have followed the recommendation of refs.~\cite{AguilarSaavedra:2018nen,DHondt:2018cww,Degrande:2020evl} and have added a factor of $1/2$ in front of the quark quadrilinear terms in the definition of the operators $\mathcal{O}^{\rm WG,1}_{QQ}$ and $\mathcal{O}^{\rm WG,8}_{QQ}$.}
\begin{align}
  \mathcal{O}^{\rm WG,1}_{QQ} ~&\equiv~ \frac{1}{2} \mathcal{Q}_{qq}^{(1)}  &\longrightarrow&& \frac{1}{2} \oLL \ ,\\
   \mathcal{O}^{\rm WG,8}_{QQ} ~&\equiv~ \frac{1}{8} \left( \mathcal{Q}_{qq}^{(3)}  + \frac{1}{3}\mathcal{Q}_{qq}^{(1)} \right) &\longrightarrow&& \frac{1}{6} \oLL \ ,
\end{align}
where we have explicitly indicated the mapping to the pure four-top operators considered in this work. The  octet operator is the same as the pure colour-octet left-handed one~\cite{DHondt:2018cww}:
\begin{align}
   \mathcal{O}^{\rm WG,8}_{QQ} ~&=~  \frac{1}{2}\bar{q}_L T^A  \gamma^{\mu} q_L   ~\bar{q}_L T^A \gamma^{\mu} q_L  \ ,
\end{align}
thus reproducing, for the full $SU(2)_{\rm L}$-invariant case, the relation found in eq.~\eqref{eq:OctetlefttoSinglet}.  The same  choices have been later adopted in refs.~\cite{Hartland:2019bjb,Degrande:2020evl}, where at LO the relation~\eqref{eq:OctetlefttoSinglet} is numerically reproduced (this is not the case at NLO as the bottom-quark interactions generate additional loop diagrams, even though it remains a good approximation). The mixed left-handed/right-handed operators as well as the pure right-handed operators match exactly our definitions.

Once the basis has been reduced, the matching between our simplified models and the EFT proceeds directly by computing and comparing the amplitudes of the four-top production process in both approaches. The case of the colour singlet is trivial as the corresponding effective operator are all included in our basis. For the colour octets, we first match the model with the corresponding octet effective operators, then use the above table to project it onto our basis. Applying a similar method for all simplified model scenarios and assuming real coupling parameters, we obtain the results shown in Table~\ref{tab:match}. An important feature is related to the vector propagators in the amplitudes, that lead to opposite signs in the large mass limit as when compared with the scalar case. This leads to the sign pattern visible from Table~\ref{tab:match}.

\section{Some technical details on event generation}
\label{sec:MC}

We generate all required four-top samples using the \amc~framework. The UFO model files for each simplified models have been made available in the \fr~model database (\href{https://feynrules.irmp.ucl.ac.be/wiki/ModelDatabaseMainPage}{https://feynrules.irmp.ucl.ac.be/wiki/ModelDatabaseMainPage}). We generate events by including all interference terms with the EW contributions to SM four-top production by typing in the \amc\ interpreter
\begin{lstlisting}[frame=single] 
  generate p p > t t~ t t~ NP^2>0 QCD=4 QED=2
\end{lstlisting}
Once the process folder has been created, we implement our choice of renormalisation and factorisation scales as a user-defined dynamical scale directly in the file \texttt{setscales.f}. We use
\begin{lstlisting}[frame=single] 
  rscale=dsqrt(max(0d0,2d0*dot(P(0,1),P(0,2))))/2
\end{lstlisting}

The CMS-TOP-18-003 analysis that we have recast is available from the \ma~Public Analysis Database at \href{https://madanalysis.irmp.ucl.ac.be/wiki/PublicAnalysisDatabase}{https://madanalysis.irmp.ucl.ac.be/wiki/PublicAnalysisDatabase} and on the {\sc MadAnalysis}~5 dataverse at \url{https://doi.org/10.14428/DVN/OFAE1G}.

\section{EFT Interference with SM electroweak contributions to four-top production}
\label{sec:EWint}

We discuss in this appendix the relevance of the SM electroweak contributions, focusing in particular on their interference with the new physics contributions. With the notable exception of colour-octet pair production (that is dominated by standard QCD interactions), new physics four-top production mechanisms mimic the SM contributions that include diagrams with Higgs-boson and $Z$-boson exchanges. Given the results of~ref.~\cite{Frederix:2017wme}, where it was shown that EW formally subleading terms at tree level give large contributions in the SM, one could wonder whether the interference of the new physics with the SM EW contributions plays an important role. Remarkably, it turns out that such an interference may actually dominate over the one with the QCD contributions, as for the production of a top-antitop pair with extra gauge bosons~\cite{Bruggisser:2021duo}.

We deconstruct these effects in the EFT case in Table~\ref{tab:cscomparison}, where the various components of the four-top production cross section are presented, and we compare our findings with those of ref.~\cite{Degrande:2020evl}.
In this table, we present the pure EFT contribution squared, at LO (first column) and NLO (fourth column), as well as its LO interference with the SM QCD (second column) and EW (third column) diagrams.  These results make use of benchmark Wilson coefficients set to $c_i/\Lambda^2 = 1$ TeV$^{-2}$. For such coefficients, the predicted cross sections are of a few fb, and the interference of the new physics diagrams with the SM electroweak ones cannot be neglected. As cross sections in this ballpark could be tested at the future high-luminosity run of the LHC (HL-LHC), we conclude that being able to rely on an C-NLO estimate of the four-top cross section including all QCD, electroweak and new physics diagrams would be certainly welcome. 
\begin{table}\begin{center}
  \setlength\tabcolsep{8pt}
  \renewcommand{\arraystretch}{1.7}
  \begin{tabular}{c|ccc|cc}
    &  \multicolumn{3}{c|}{LO}  & \multicolumn{2}{c}{NLO}\\
	& $\big|{\rm new~physics}\big|^2$ & Int. QCD only & Int. EW only& QCD~\cite{Degrande:2020evl} & via $K_{\rm SM}$	\\
		\hline 	$\oLL/2$  	& $0.8\pmerr{44}{28}$ fb& $0.20\pmerr{47}{31}$ fb& $-0.80\pmerr{41}{28}$ fb & $1.6\pmerr{3}{10}$ fb& $0.62\pmerr{18}{22}$ fb \\
	 $\oLR$  	& $1.1\pmerr{45}{27}$ fb& $-0.02\pmerr{32}{16}$ fb& $0.60\pmerr{44}{28}$fb& $1.84\pmerr{3}{10}$ fb& $3.9\pmerr{21}{26}$ fb \\
	 $\oRR$  	& $3.4\pmerr{44}{28}$ fb& $0.39\pmerr{55}{29}$ fb& $-1.42\pmerr{40}{30}$ fb& $6.14\pmerr{3}{10}$ fb& $5.5\pmerr{20}{22}$  fb\\
	  $\oTaLR$  	& $0.28\pmerr{44}{29}$ fb& $0.22\pmerr{52}{35}$ fb& $-0.49\pmerr{42}{28}$ fb& $0.69\pmerr{3}{8}$ fb& $0.01^{+0.10}_{-0.04}$  fb\\ \hline
	SM  	& / & $4.7\pmerr{66}{38}$ fb& $0.50^{+0.95}_{-0.87}$ fb&  / & $11.97\pmerr{18}{21}$  fb
		\end{tabular}
		\caption{Different new physics components of the $pp\to t\bar t t\bar t$ cross section at the LHC, for $\sqrt{S}=13$~TeV and in the EFT approach. We consider various operators $\mathcal{O}_i$ with which we associate a Wilson coefficient $c_i/\Lambda^2 = 1$ TeV$^{-2}$. The first three columns represent the LO contributions (the new physics squared piece, the interference of the new physics diagrams with the SM QCD diagrams, and the one with the electroweak diagrams), the fourth column the new physics contributions at NLO in $\alpha_S$ and the last one the results obtained by using the scheme of eq.~\eqref{eq:multscheme}. In the last line of the table, we show the SM LO results by considering only the QCD diagrams, and by including both the QCD and EW contributions, as well as the full NLO result~\cite{Frederix:2017wme}.}
		\label{tab:cscomparison}
	\end{center}
\end{table}%

\bibliographystyle{JHEP}
\bibliography{biblio}

\end{document}